\documentclass[pdflatex,sn-nature]{sn-jnl}% Style for submissions to Nature Portfolio journals
\usepackage{graphicx}%
\usepackage{multirow}%
\usepackage{amsmath,amssymb,amsfonts}%
\usepackage{amsthm}%
\usepackage{mathrsfs}%
\usepackage[title]{appendix}%
\usepackage{xcolor}%
\usepackage{textcomp}%
\usepackage{manyfoot}%
\usepackage{booktabs}%
\usepackage{algorithm}%
\usepackage{algorithmicx}%
\usepackage{algpseudocode}%
\usepackage{listings}%
\usepackage{lipsum} % generates filler text

\theoremstyle{thmstyleone}%
\theoremstyle{thmstyletwo}%

\theoremstyle{thmstylethree}%

\begin{document}

%\title[Article Title]{Transformer Based Learning for Magnetic Field and Ocean Characterization at Europa} %SR
%\title[Article Title]{Revealing ocean characteristics at Jupiter's moon Europa using Transformer-based machine learning} %TN 
%\title[Article Title]{Transformers help characterize Europa’s ocean through magnetic field inference} %SR
%\title[Article Title]{Rapid Magnetic Field Inference For Ocean %Characterization at Europa using Transformer-Based Learning} %SR
%\title[Article Title]{LEAP: Rapid Magnetic Field Inference For Ocean Characterization at Europa} %SR
\title[Article Title]{LEAP: A Rapid Neural Surrogate of Multi-Fluid MHD at Europa} %SR


\author*[1]{\fnm{Sachin A.} \sur{Reddy}}\email{sachin.reddy@jpl.nasa.gov}

\author[2, 3, 4]{\fnm{Abigail R.} \sur{Azari}}\email{aazari@ualberta.ca}

\author[1]{\fnm{Corey J.} \sur{Cochrane}}\email{corey.cochrane@jpl.nasa.gov}

\author[5]{\fnm{Xianzhe} \sur{Jia}}\email{xzjia@umich.edu}

\author[6]{\fnm{Tom A.} \sur{Nordheim}}\email{tom.nordheim@jhuapl.edu}

\author[1]{\fnm{Lukas} \sur{Mandrake}}\email{lukas.mandrake@jpl.nasa.gov}

\author[1]{\fnm{Steven D.} \sur{Vance}}\email{steven.d.vance@jpl.nasa.gov}

\author[7]{\fnm{Camilla} \sur{Harris}}\email{camilla.harris@ucl.ac.uk}

\author[8]{\fnm{Ioana} \sur{Ciuc\u{a}}}\email{iciuca@stanford.edu}

\affil*[1]{\orgdiv{Jet Propulsion Laboratory}, \orgname{California Institute of Technology}, \orgaddress{\street{4800 Oak Grove Drive}, \city{Pasadena}, \postcode{91109}, \state{CA}, \country{United States}}}

\affil[2]{\orgdiv{Physics Department}, \orgname{University of Alberta}, \orgaddress{\street{1011 - 88 Avenue}, \city{Edmonton}, \postcode{T6G 2G5}, \state{Alberta}, \country{Canada}}}

\affil[3]{\orgdiv{Electrical and Computer Engineering Department}, \orgname{University of Alberta}, \orgaddress{\street{1011 - 88 Avenue}, \city{Edmonton}, \postcode{T6G 2G5}, \state{Alberta}, \country{Canada}}}

\affil[4]{\orgdiv{Canada CIFAR AI Chair}, \orgname{Alberta Machine Intelligence Institute}, \orgaddress{\street{10065 Jasper Ave 1101}, \city{Edmonton}, \postcode{T5J 1S5}, \state{Alberta}, \country{Canada}}}

\affil[5]{\orgdiv{Department of Climate and Space Sciences and Engineering}, \orgname{University of Michigan}, \orgaddress{\street{500 S State St}, \city{Michigan}, \postcode{48109}, \state{MI}, \country{USA}}}

\affil[6]{\orgname{Johns Hopkins Applied Physics Laboratory}, \orgaddress{\street{11100 Johns Hopkins Rd}, \city{Laurel}, \postcode{20723}, \state{MD}, \country{Country}}}

\affil[7]{\orgdiv{Advanced Research Computing Centre}, \orgname{University College London}, \orgaddress{\street{Gower Street}, \city{London}, \postcode{WC1E 6BT}, \country{UK}}}

\affil[8]{\orgdiv{Kavli Institute for Particle Astrophysics and Cosmology}, \orgname{Stanford University}, \orgaddress{\street{452 Lomita Mall}, \city{Stanford}, \postcode{94305}, \state{California}, \country{United States}}}


%MHD First (2026-02-12
%\abstract{High fidelity computational models (e.g. magnetohydrodynamic, MHD) are essential for resolving the perturbations that arise from Jupiter's plasma interaction with Europa. These perturbations complicate the measurement of Europa's ocean via magnetic induction, which is a key objective of the Europa Clipper and JUICE missions in the search for life beyond Earth. 
%Though comprehensive, the computational cost of MHD codes, limits our ability to fully constrain Europa's habitability. To address this, we introduce \textbf{Learning Europa's Atmosphere and Plasma (LEAP)}, a transformer-surrogate that predicts magnetic fields using outputs from a state-of-the-art multi-fluid MHD simulation as training data. LEAP makes predictions along spacecraft trajectories and evaluates in milliseconds on a laptop while maintaining predictions within $\pm$ 3 nT. For the Galileo $E4$ and $E14$ flybys of Europa, LEAP closes matches the MHD parent model in accuracy, while being $\sim$40,000x faster than the physics-only code. Its enhanced speed enables parameter surveys and probabilistic estimations of plasma conditions, demonstrating how machine learning can facilitate novel scientific discovery and improve the ocean characterization process. LEAP can also inform future MHD simulations while also learning from them. This framework could be expanded to planning future missions or to other high priority bodies, including Uranus and Neptune.}

%Europa First
\abstract{Characterizing Europa’s subsurface ocean is a key objective of the Europa Clipper and JUICE missions in the search for life beyond Earth. Although the ocean’s induced magnetic field provides key constraints on habitability, interpretation is complicated by perturbations arising from Jupiter's plasma interaction with Europa. Physics-based models (e.g. magnetohydrodynamic, MHD) required to characterize these effects are physically comprehensive, but have a prohibitive computational cost. To address this, we introduce \textbf{Learning Europa's Atmosphere and Plasma (LEAP)}, a transformer-based surrogate trained on outputs from a state-of-the-art multi-fluid MHD code to predict magnetic field perturbations along spacecraft trajectories. LEAP evaluates in milliseconds on a laptop, whereas MHD takes 12 hrs on a high-performance computer ($\sim$40,000$\times$ speed-up). The model has test set errors of $\pm$ 2.6 nT, and for the Galileo \textit{E4} and \textit{E14} flybys of Europa it matches the parent MHD model in accuracy. Its enhanced speed enables large-scale parameter surveys and probabilistic estimations of plasma conditions, establishing a new framework for accelerated plasma interaction modeling. LEAP can also inform future MHD simulations while learning from them. Beyond Europa, this framework could be expanded to planning future missions or to other high-priority bodies, including Uranus and Neptune.}

\keywords{Europa, Magnetic Fields, Ocean World Characterization, Transformers, Habitable Worlds}

\maketitle

\section{Introduction}\label{sec-intro}
%750 words
%Europa and habitability
Jupiter's moon Europa is a prime target in the search for life \cite{vance2023investigating}, and will be visited by NASA's Europa Clipper and ESA's JUICE missions starting in 2031 \cite{Grasset2013, pappalardo2024science}. These robotic missions will investigate Europa's habitability by, in part, characterizing its subsurface ocean \cite{pappalardo2024science}. The first major evidence of an ocean came from the Galileo mission (1996-2003) when its magnetometer measured perturbations in the vicinity of Europa that were best explained by an induced magnetic field arising from a conductive layer beneath Europa's ice-covered surface \cite{khurana1998induced, kivelson2000galileo, zimmer2000subsurface}. Since ice is a poor conductor, the measurements implied the presence of a global salt water ocean that responds inductively to an external time-varying field \cite{jia2010magnetic}. %The ocean hypothesis was further corroborated by the paucity of impact features, subdued topography, and expected tidal heating of Europa \cite{ross1987tidal, pappalardo1999does, bierhaus2001pwyll}.
Magnetic induction provides an indirect link to habitability: the observed signal is related to ocean conductivity, which is reflective of the salt content, which in turn tells us about the types of life Europa's ocean might support \cite{vance2023investigating}. 

Europa orbits deep within Jupiter’s magnetosphere ($\sim$9~$\mathrm{R_J}$), where it continuously interacts with the planet’s magnetic field and co-rotating plasma disk \cite{jia2010magnetic}. Because Jupiter’s magnetic dipole is tilted by roughly 10° relative to its rotational axis, the plasma sheet oscillates across the ecliptic plane over Jupiter’s 11.2-hr synodic period \cite{kivelson2004magnetospheric}. As a result, Europa encounters a highly time-variable plasma environment that is modulated by both this synodic variation and its own 85.2-hr orbital period. Owing to the frozen-in-flux condition, the plasma and background magnetic field largely move together, and are strongly perturbed as they interact with Europa. This interaction produces semi-deterministic electromagnetic features, including magnetic pile-up on the trailing hemisphere, upstream field-line stretching, and Alfvén wings extending to the north and south of the moon \cite{neubauer1999alfven, volwerk2007europa, jia2010magnetic, harris2021multi}. These plasma-driven perturbations generate additional time-varying magnetic fields which are distinct from Jupiter’s background field and are referred to as plasma interaction fields (PIFs) \cite{cochrane2025stronger}. Therefore, the magnetic field measured near Europa reflects the superposition of Jupiter’s background field, plasma interaction fields, and the subsurface ocean’s inductive response to both. Because PIFs can be comparable in magnitude to the inducing Jovian field and vary rapidly in space and time \cite{neubauer1980nonlinear, neubauer1999alfven, harris2021multi, saur2010induced}, they must be characterized with sufficient fidelity to properly isolate the oceanic induction signal needed for habitability assessments \cite{vance2023investigating, kivelson2023ecm}. These electromagnetic interaction features are complex and require resource-intensive computational models, such as magnetohydrodynamic (MHD) codes \cite{harris2021multi}, to properly interpret in-situ magnetic field measurements \cite{jia2010magnetic, kivelson2023ecm}. %On Europa Clipper, the magnetic field will be measured by the Europa Clipper Magnetometer (ECM) \cite{kivelson2023ecm}, while plasma properties will be constrained by the Plasma Instrument for Magnetic Sounding (PIMS) \cite{westlake2023plasma}.
%Eddy currents are just loops of electric current that appear in a conductor when the magnetic field passing through it changes with time.

%Modeling efforts
Over the years, numerous MHD models have been developed to characterize plasma interactions between Europa and Jupiter. These models have evolved from one ion fluid descriptions \cite{kabin1999europa, schilling2008influence}, to two \cite{rubin2015self, jia2018evidence}, and finally, to the state-of-the-art, three-ion-fluid models \cite{harris2021multi, harris2022multi}. These three-fluid models have been validated against Galileo flybys $E4$ and $E14$ \cite{harris2021multi}, and have been used to explore the role of mass density, atmospheric scale height, and neutral density in shaping Europa’s electromagnetic environment \cite{harris2022multi}. A MHD model was also used to infer the presence of plume activity at Europa \cite{jia2018evidence}. More recently, a single-fluid MHD model has been augmented to include electron beam physics, yielding improved agreement with Juno flyby observations of Europa \cite{allegrini2024electron, cervantes2025mhd}. In parallel with fluid-based approaches, hybrid fluid--kinetic models have been developed to capture kinetic effects absent from MHD descriptions \cite{muller2011aikef, addison2024magnetic}, with additional examples summarized therein.

%Issues with current models
Despite their physical fidelity, high-resolution MHD models are computationally expensive to run. They typically require more than 12 hours to complete on a 2,000-core high-performance computer (HPC). % such as Pleiades \cite{biswas2017pleiades}. 
Matching observations to simulations often entails multiple runs, resulting in a potentially days- or weeks-long process. Europa Clipper will return data from close flybys of Europa every 2--3 weeks over its nominal mission duration \cite{pappalardo2024science}, representing a bottleneck in real-time assessment in the ocean characterization process as the tour progresses. Moreover, the high computational cost precludes systematic parameter surveys (forward modeling) and robust parameter estimation via inverse or simulation-based inference \cite{Cranmer2020}, both of which are essential given the limited observational constraints on Europa’s plasma environment.

%ML-MHD-Emulator
Machine learning (ML) emulators provide a way forward by approximating the behavior of computationally expensive models such as MHD. Transformer architectures \cite{vaswani2017attention}, which leverage attention to capture long-range dependencies in sequential data, have recently been adapted for modeling dynamical and PDE-governed physical systems \cite{holzschuh2025transformer, wan2025pesanet}. Within planetary science, an MHD model of the Ganymede system \cite{duling2022ganymede} has been emulated using a Bayesian surrogate approach \cite{azari2023simulation}. In space weather applications, emulators based on echo-state networks have been developed for MHD simulations of Earth’s polar ionosphere \cite{tanaka1994finite, kataoka2024machine}, while graph neural networks have been applied to hybrid kinetic--MHD models of the terrestrial magnetosphere \cite{holmberg2025graph, von2014vlasiator}. In astrophysics, curated MHD datasets of interstellar turbulence have been released to enable the development of surrogate models \cite{burkhart2020catalogue}.

\begin{figure}[h]
    \centering
    \includegraphics[width=1\textwidth]{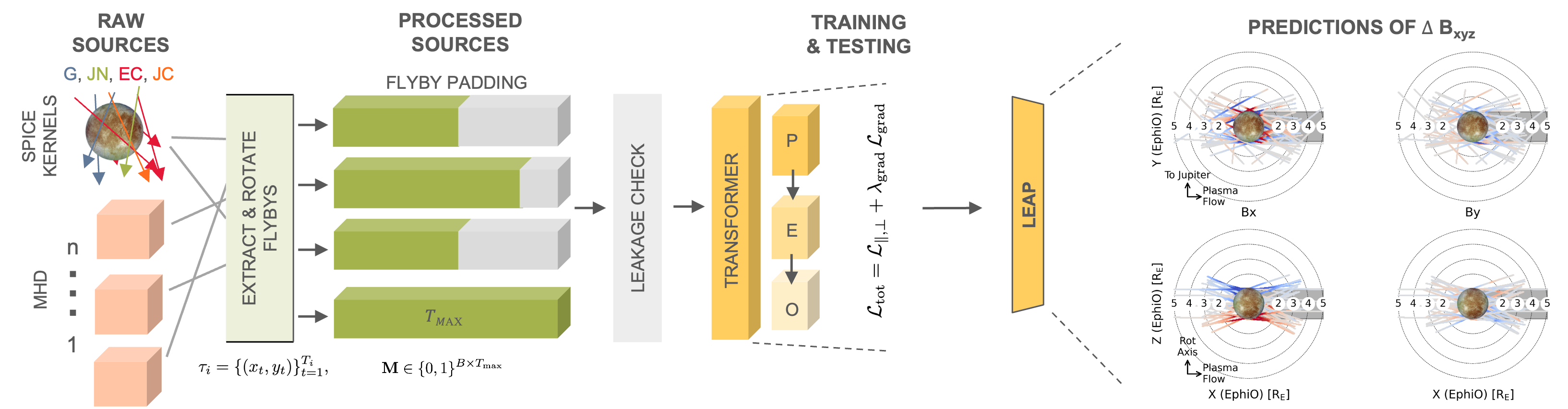}
    \caption{\textbf{The development of the Learning Europa's Atmosphere and Plasma (LEAP) model.} LEAP ingests MHD data and flyby trajectories, rotating and padding them before performing a rigorous leakage assessment and feeding them into a lightweight transformer regressor. \texttt{SPICE} kernels mission key: G = Galileo, JN = Juno, EC = Europa Clipper and JC = JUICE.} 
    %\caption{\textbf{The development of LEAP and the ocean characterization pipeline.} \textbf{a,} LEAP ingests MHD data and flyby trajectories, rotating and padding them before feeding them into a lightweight transformer regressor. The transformer then acts as likelihood model for a Bayesian inversion (via Markov-chain Monte Carlo) estimator. LEAP enables novel parameter estimates, surveys, and importances. Spice kernels mission key: G = Galileo, JN = Juno, EC = Europa Clipper and JC = JUICE. \textbf{b,} the ocean characterization pipeline with respect to the upcoming missions, MHD code (12 hr) and LEAP ($<$1 s). \textit{PlanetProfile} is an open-source self-consistent interior structure toolkit for modeling ocean worlds and rocky dwarf planets \cite{vance2018geophysical,styczinski2023planetprofile}.}
    \label{fig1}
\end{figure}

%In this study
In this study, we introduce \textbf{Learning Europa's Atmosphere and Plasma (LEAP)}: a transformer-based surrogate that rapidly predicts magnetic field perturbations along spacecraft trajectories. LEAP can be configured to any spacecraft flyby configuration (e.g. JUICE and Europa Clipper) and when combined with a background field model (e.g., Khurana \cite{khurana1997euler}), can predict the total expected field for a given past or future flyby. In the next section, we describe the performance of LEAP relative to its parent MHD model. We then demonstrate novel analyses of the Galileo \textit{E4} and \textit{E14}, and JUICE \textit{1E7} and \textit{2E8} flybys enabled by its rapid evaluation time (1 s). The LEAP framework is summarized in Figure \ref{fig1} with further technical information in the Methods section. 

%---------------------------------------------------------------------------------------
%---------------------------------------------------------------------------------------
%---------------------------------------------------------------------------------------
%----------------------------------  R E S U L T S  -------------------------------
%---------------------------------------------------------------------------------------
%---------------------------------------------------------------------------------------
%---------------------------------------------------------------------------------------
%---------------------------------------------------------------------------------------
%1375 words
\section{Results}\label{sec-results}
%Results into (do last)
%In this section, we validate the performance and utility of the model. We first compare it to the MHD test set and then to the Galileo magnetometer observations. Then, we apply LEAP to the upcoming JUICE flybys of Europa, comparing 1000's of model runs to the expected physics of the system. Next, we treat the transformer as a forward model and present a novel re-estimation of the MHD plasma parameters at Europa using a Bayesian inversion method via Markov-chain Monte Carlo. Finally, we assess which parameters mainly control Europa's simulated magnetic field environment; helping to guide future MHD runs and to generate more data from which LEAP can learn.

\begin{figure}[htp]
    \centering
    \includegraphics[width=0.95\textwidth]{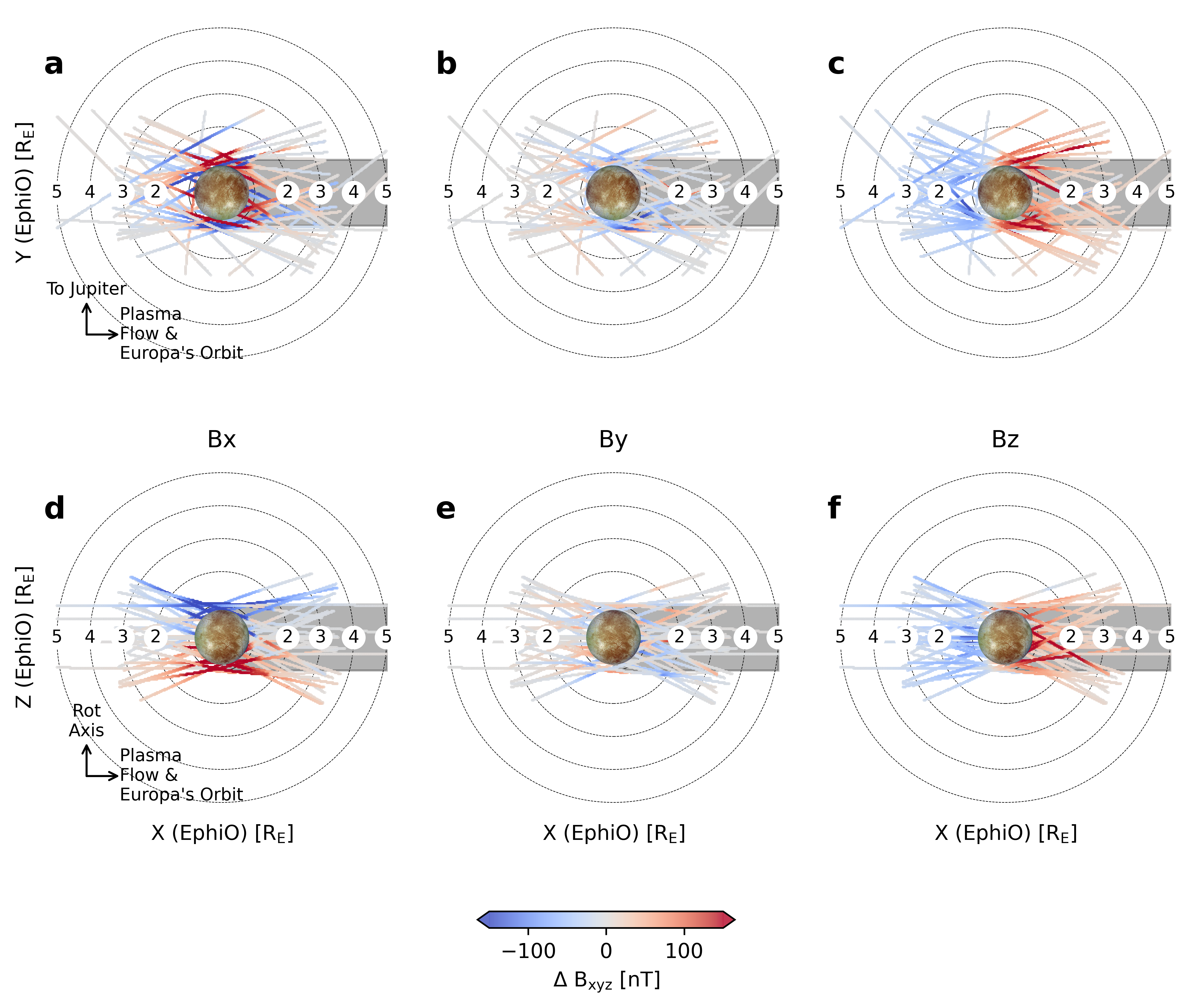}
    \caption{\textbf{LEAP predictions of the magnetic field along select flyby trajectories from within the test set}}
    \label{fig2}
\end{figure}

%---------------------------------------------------------------------------------------
%Predictions on Figure 2
In Figure \ref{fig2} we show LEAP predictions of $\Delta B_{xyz}$ along real flybys from within the test set in the EPhiO coordinate system. Across Fig. \ref{fig2} we see strong perturbations close to the moon and less farther away, especially in $B_x$ and $B_z$. In Figs \ref{fig2}c and \ref{fig2}f, magnetic pile-up in the trailing hemisphere and stretching in the leading hemisphere can be observed, as denoted by the shift in $\Delta B_z$ from roughly -100 nT (upstream) to +100 nT (downstream). In Fig. \ref{fig2}d, Alfvén wings \cite{neubauer1999alfven, volwerk2007europa} are correctly predicted, as denoted by the flipped polarity of the $B_x$ component north and south of Europa. These results are promising and demonstrate that LEAP has the fidelity to learn the expected physics of the system \cite{kivelson2004magnetospheric, jia2010magnetic, harris2021multi}, which itself is captured by the parent MHD model.

\begin{figure}[htp]
    \centering
    \includegraphics[width=0.95\textwidth]{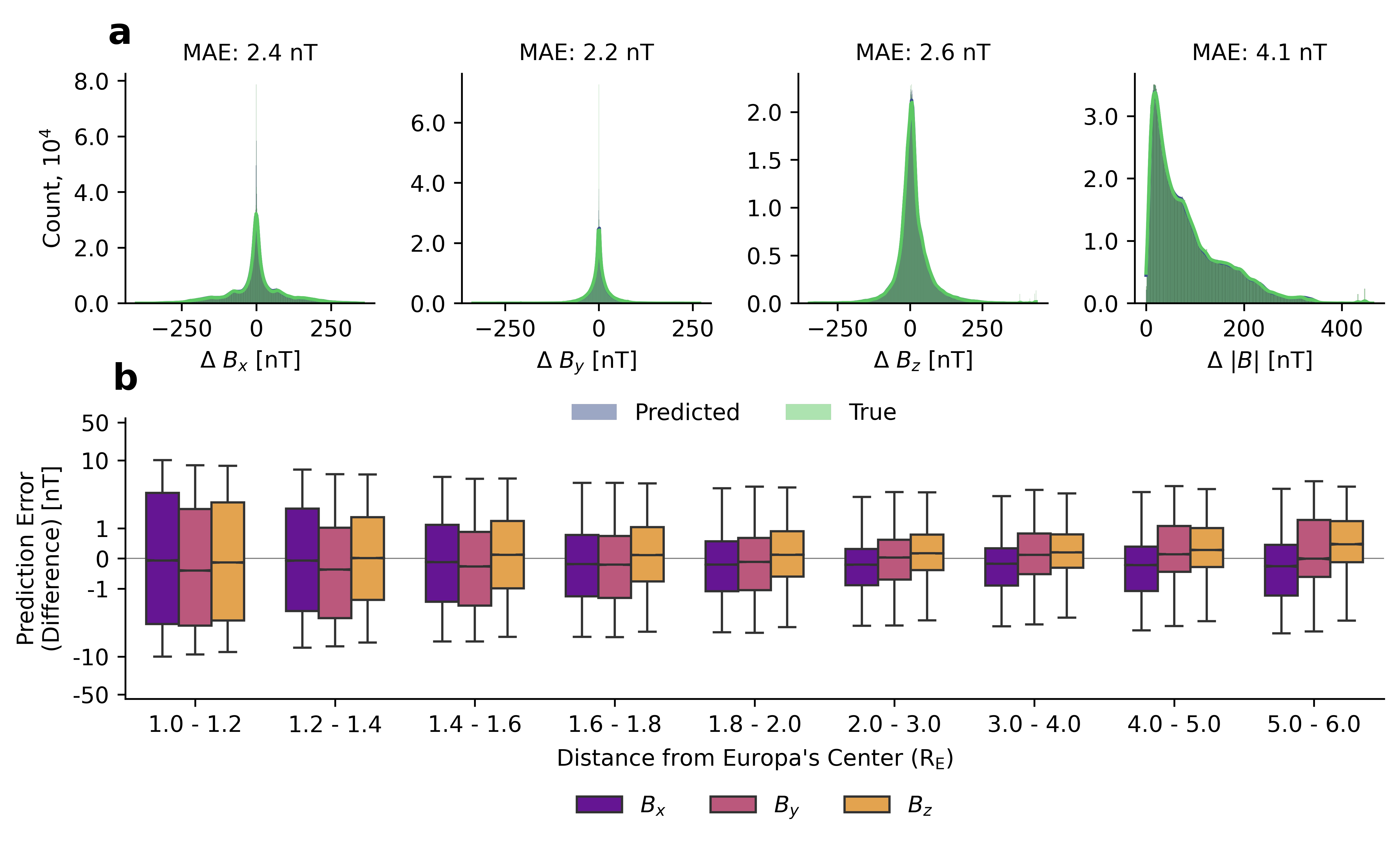}
    \caption{\textbf{Performance of LEAP on the MHD test set.} \textbf{a,} Mean absolute error of LEAP across the MHD target range. \textbf{b,} prediction errors of $B_x$, $B_y$, $B_z$ as a function of radial distance from Europa}
    \label{fig3}
\end{figure}

%---------------------
%Figure 3
%MAE between ML and MHD (Fig 2a)
Figure \ref{fig3}a shows the performance of LEAP with respect to the true MHD values, computed over all flybys ($\sum T$). All component errors have a mean absolute error (MAE) of $< 2.7$~nT, with no obvious bias across the target range. For comparison, the ECM investigation -- whose data the MHD code and LEAP will compare to starting in 2031 -- requires the amplitude of the 11.2~hr synodic and 85.1~hr orbital periods to be estimated to within a precision of 1.5~nT \cite{kivelson2023ecm}. Note that $\Delta|B|$ is not explicitly trained but derived from $|B| = \sqrt{(B_x^2 + B_y^2 + B_z^2)}$.

%Difference Error (Fig 2b)
Figure \ref{fig3}b shows prediction error versus radial distance from Europa. All predictions have a mean error of $\pm$1~nT and remain within 10 nT up to the error bar whiskers (1.5$\times$ inter-quartile range). Fig. \ref{fig3}b also shows that the prediction error improves with radial distance from the moon. This is expected as the environment beyond 3$R_E$ is less dynamic than the region close to Europa (see Fig. \ref{fig2}), resulting in a more steady-state and consistent target value. Although training objectives are typically designed to prioritize labels with the greatest error, we found that adopting a vanilla L1 or L2 loss led to an even greater error near Europa. We addressed this, in part, with the radial weighting term $w$ that forms part of our anisotropy loss function (see Eqs. \ref{eq:training-obj-radial}-\ref{eq:training-obj-par-perp}).

\begin{figure}[htp]
    \centering
    \includegraphics[width=\textwidth]{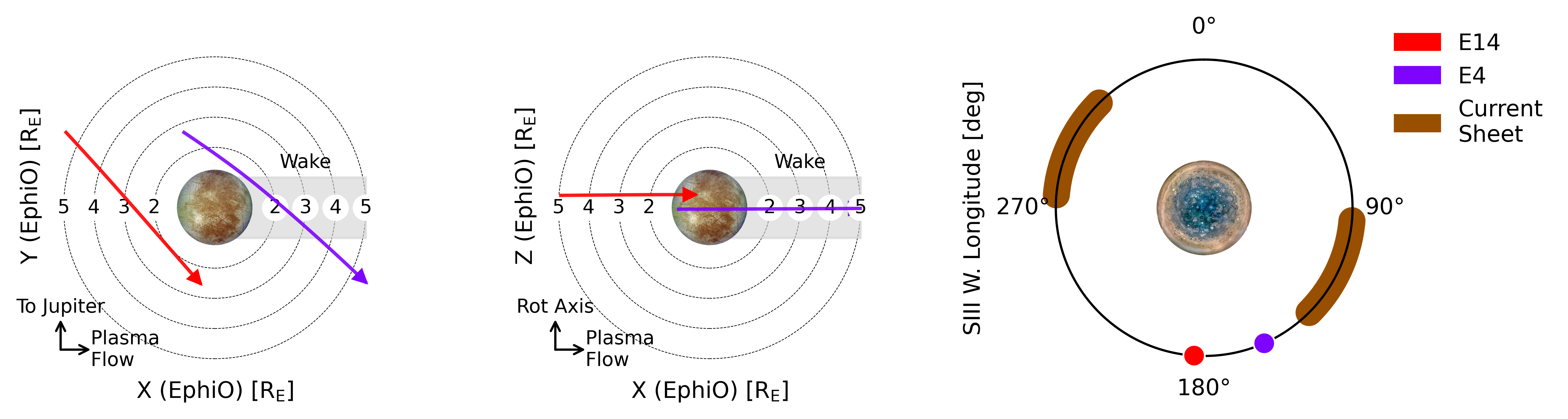}
    \caption{\textbf{The Galileo \textit{E4} and \textit{E14} flybys trajectories}. The gray region represents Europa's geometric wake. Data is in the EPhiO coordinate system.}
    \label{fig4}
\end{figure}

%---------------------
%Figure 4
%$E4$ and $E14$ Geometry
Next, we use LEAP to predict the magnetic field along the Galileo \textit{E4} and \textit{E14} flybys of Europa, comparing it to previously-run MHD simulations \cite{harris2021multi}, which are considered state-of-the-art. Figure \ref{fig4} shows the geometry of the flyby trajectories with respect to Europa and its geometric wake, Jupiter, and the Jovian current sheet. 

\begin{figure}[htp]
    \centering
    \includegraphics[width=0.95\textwidth]{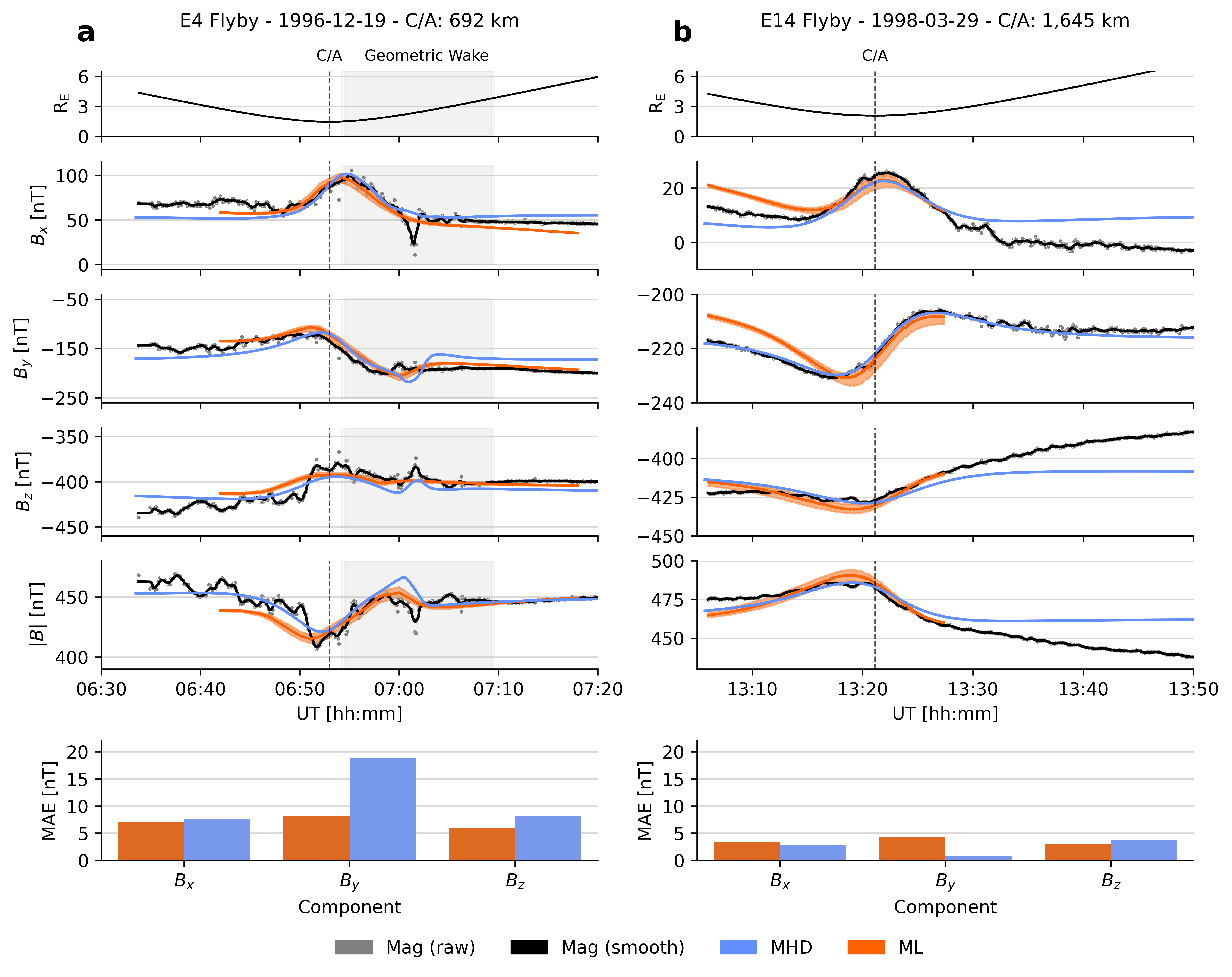}
    \caption{\textbf{Performance of LEAP with respect to the Galileo $E4$ and $E14$ flybys and MHD simulations}. The gray dashed-line denotes the closest approach, and the gray box denotes the geometric wake. The bar charts compare the MHD and LEAP to the observations using a root mean square error. Data is in the EPhiO coordinate system.}
    \label{fig5}
\end{figure}

%---------------------
%Figure 5
%$E4$ and $E14$ Performance
Figure \ref{fig5} shows the real observational data, the MHD runs, and LEAP predictions. These flybys and the associated data were held out from the training set and have not been seen by LEAP before. As LEAP evaluates near-instantaneously ($<$ 1 sec), this allows us to run a large-scale parameter sweep of select model inputs. In this instance we chose to vary the magnetospheric O$^+$ mass density ($\rho_0$), Europa's atmospheric surface density ($n_0$), and Europa's atmospheric scale height ($H_0$) within their interpolation range. To enable a comparison with the MHD and flybys, we add the Khurana Jovian background field model \cite{khurana1997euler} to the LEAP predictions, plotting $\Delta \hat{B}_{xyz} + B_{J, xyz}$. We survey 10,000 potential conditions per flyby and reject predictions where $|\mathbf{\hat{B}}_{E4}| >$ 8.5 nT and $|\mathbf{\hat{B}}_{E14}| >$ 5.5 nT, resulting 171 and 573 acceptable predictions, respectively. Next, we calculate the 5th and 95th percentile bounds, with the median (50th percentile) shown in dark orange. The raw observations (gray) are smoothed using a 60-second rolling window (black) to provide a reasonable comparison. The light gray shaded area in Figure \ref{fig5}a represents the geometric wake -- the downstream region influenced by the moon’s obstruction of the Jovian magnetospheric plasma flow \cite{kivelson2004magnetospheric, jia2010magnetic}. Closest approach (C/A) is denoted by the dashed vertical line. All data are in the EPhiO coordinate system. Finally, the bar charts in the figure below summarize the mean absolute error (MAE) between each method relative to the Galileo observations. 

For the \textit{E4} flyby (Fig. \ref{fig5}a), LEAP matches the parent MHD model in accuracy and the parameter survey allows for statistical improvements in all three components, especially in $B_y$. This result is significant because Europa's induction response ($B_{ocean}$ in Eq. \ref{eq:B_obs}) is predominantly observed in the $B_x$ and $B_y$ components \cite{kivelson2004magnetospheric,harris2022multi}. This means that LEAP, alongside the MHD, could be used as part of the induction experiment \cite{pappalardo2024science, kivelson2023ecm} to characterize Europa's subsurface ocean, should training data with a variable induction response become available. Both LEAP and the MHD underperform at 07:01~UT, where a localized field variation is not captured by either model in $B_x$ or $B_z$, though LEAP deviates less in $B_y$. For \textit{E14} (Fig. \ref{fig5}b), the ML model captures the magnitude and overall structure of the magnetic field perturbations ($\Delta B$), especially at closest approach. The deviations in $B_x$ and $B_y$ between 13:05~UT and 13:15~UT arise from the mismatch in coordinate systems between the spacecraft, MHD and LEAP (\texttt{SPICE} kernels), which cannot be avoided at this stage. This same artifact also affects the MHD beyond 13:30~UT in $B_x$ and $B_z$, and in both cases, these divergences are not reflective of a poor prediction. This may be investigated further in future works. Finally, the LEAP prediction is capped at 13:27~UT as this is the boundary of the training volume ($\pm$ 2.5 $R_E$ in the $y$ or Jupiter direction).
%---------------------------------------------------------------------------------------
\subsection{Parameter Estimation}
%intro
Constraining the plasma and neutral conditions at Europa is essential for the upcoming missions. Within the LEAP framework, we formulate this as an inverse problem using Simulation-Based Inference (SBI) \cite{Cranmer2020}. The transformer's $\sim$40{,}000$\times$ acceleration enables the trained emulator to function as a near-instantaneous probabilistic forward model (Eqs.~\ref{eq:likelihood}–\ref{eq:Bayes}), allowing estimation of selected parameters from the parent MHD model \cite{harris2021multi}. We emphasize that the inversion is performed on \textit{simulated} MHD outputs rather than spacecraft observations. For consistency, we invert the same parameters varied in Figure~\ref{fig5}: $\rho_0$, $n_0$, and $H_0$.

\begin{figure}[h]
    \centering
    \includegraphics[width=0.9\textwidth]{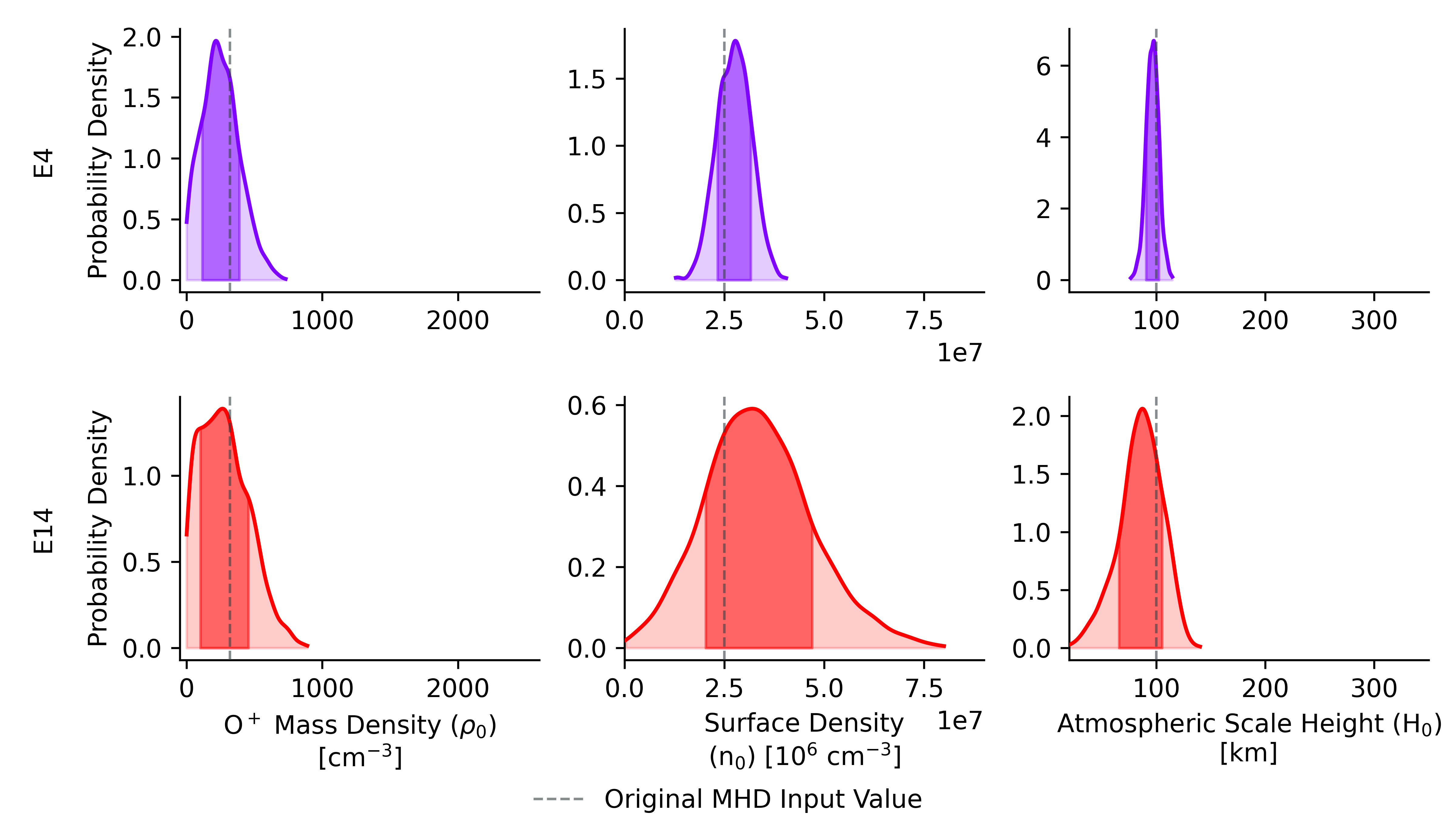}
    \caption{\textbf{Estimating plasma conditions based on \textit{simulated} flyby data via SBI}. The shaded region represents parameter values within $\pm \sigma$ of the posterior mean.}
    \label{fig6}
\end{figure}

%figure description
Figure~\ref{fig6} presents the posterior distributions obtained from simulated MHD data along the \textit{E4} and \textit{E14} flybys using a Bayesian approach (Eqs.~\ref{eq:likelihood}–\ref{eq:Bayes}). The dashed vertical line denotes the original MHD input value used in Fig.~\ref{fig5} and \cite{harris2022multi}. The shaded region indicates the range of parameter values within $\pm \sigma$ of the posterior mean. Additional details are provided in the Methods and Table~\ref{tab3}.
%analysis
For \textit{E4}, the most probable values are only slightly offset from the MHD input values for all three parameters. The posteriors of $\rho_0$ and $n_0$ show moderate spread, whereas $H_0$ is comparatively well constrained. In \textit{E14}, a similar offset between the MHD input and posterior mean is observed, but all three parameters span a broader region of parameter space, particularly $n_0$. The posterior mean surface density for \textit{E14} is not substantially higher than that of \textit{E4}; instead, the primary distinction lies in the increased width of the $n_0$ distribution. This broader admissible range may reflect weaker constraint or enhanced variability in the trailing hemisphere plasma environment. Although we do not recover the order-of-magnitude density enhancement reported from radio occultation measurements \cite{kliore1997ionosphere}, the expanded posterior for $n_0$ is qualitatively consistent with a more variable or structured plasma regime on the trailing hemisphere. Overall, these simulated results demonstrate that surrogates such as LEAP can provide probabilistic constraints on model parameters ($\theta$), offering an significant advantage over single-point deterministic estimates.

%---------------------------------------------------------------------------------------
\subsection{Parameter Surveys}
%plot of JUICE flybys estimations
In addition to estimating conditions for past flybys, we also apply LEAP to future and unknown conditions. We choose the two JUICE flybys of Europa as they have favorable trajectory geometries for validating the predictions. We perform a parameter survey of 4,000 random values within the interpolation range of the same three parameters: upstream magnetospheric O$^+$ mass density ($\rho_0$), Europa's atmospheric surface density ($n_0$), and Europa's atmospheric scale height ($H_0$). The remaining inputs for the model are fixed and include the coordinates of the trajectory (x, y, z, $R_E$) in the EPhiO coordinate system and are extracted from the latest tour design (\texttt{juice\_crema\_5\_1\_150lb\_23\_1\_a3\_2\_v01}) \cite{boutonnet2024designing, masters2025magnetosphere} using \texttt{SPICE} kernels and the Python package \texttt{SpiceyPy}. The Jovian magnetic field ($B_{jx}, B_{jy}, B_{jz}$) at Europa's orbit is calculated using the Khurana model \cite{khurana1997euler}. These values form both the inputs and, as before, are added to the predictions to make them more realistic e.g., $ \hat{B}_{xyz} =\Delta \hat{B}_{xyz} + B_{J, xyz}$. The induced dipole ($M$) follows the form of $M = \sqrt{M_x^2+ M_y^2}$, where $M_x = -B_{jx}/2, \ M_y = -B_{jx}/2$ in alignment with the parent MHD code \cite{harris2021multi, harris2022multi}. Owing to training data restrictions, all predictions assume a 100\% ocean induction efficiency using the analytical model described above. Future versions of LEAP may be trained on MHD outputs with a variable $M$ response or could be integrated with more advanced ocean models such as PlanetProfile \cite{styczinski2023planetprofile}, which in turn would produce different predictions for the JUICE flybys shown herein.

\begin{figure}[h]
    \centering
    \includegraphics[width=1\textwidth]{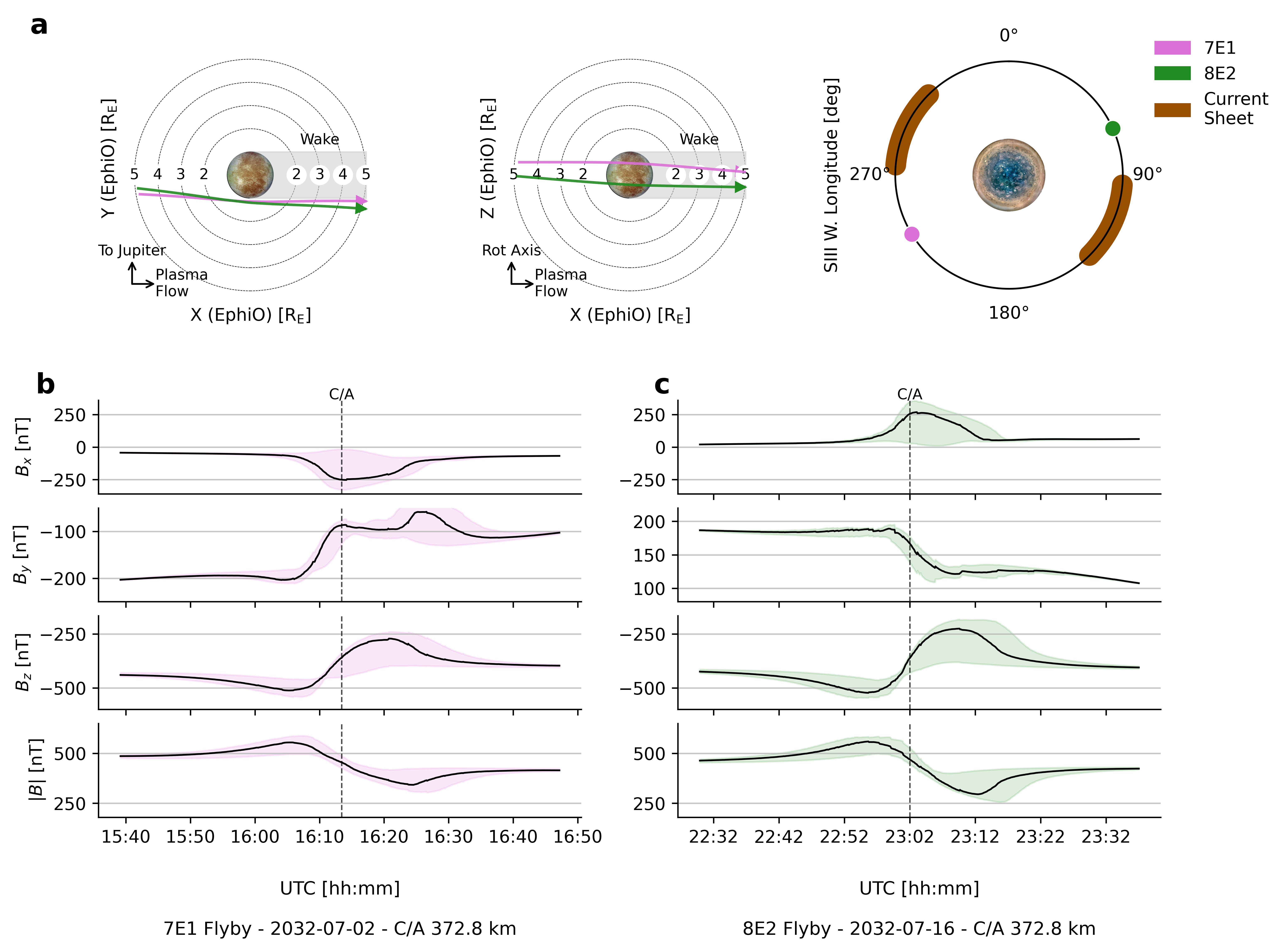}
    \caption{\textbf{Predicting magnetic field components for the JUICE flybys of Europa.} \textbf{a,} The expected trajectory of JUICE and the positions of the spacecraft and Europa with respect to Jupiter and its current sheet. \textbf{b,c}  We plot $\hat{\mathbf{B}}_{xyz} =\Delta \hat{\mathbf{B}}_{xyz}$ (LEAP) + $\mathbf{B}_{J, xyz}$ (Khurana~\cite{khurana1997euler}) for the two flybys with respect to UTC and closest approach. Data is in the EPhiO coordinate system.}
    \label{fig7}
\end{figure}

Figure \ref{fig7}a shows the different flyby configurations of 7E1 and 8E2 with respect to Europa and the Jovian current sheet. As with Figure \ref{fig5}, we plot the 5th and 95th percentile bounds, with the median (50th percentile) shown in black. Fortuitously, these flybys share similar x-y geometries, but vary in z: 7E1 approaches Europa in its northern hemisphere while 8E2 crosses its southern. This allows us to clearly observe both the Alfvén wing structures in $B_x$ \cite{neubauer1999alfven, volwerk2007europa}, and magnetic pile-up and stretching in $B_z$ \cite{kivelson2004magnetospheric}, akin to the results in Figure \ref{fig2}c-d,f. Figure \ref{fig7}b-c also shows that the variability of the prediction tends to 0 as we move away from C/A; this is expected as only the background model \cite{khurana1997euler} should be present here. Once again, the prediction of these features suggests that LEAP is capturing the expected electrodynamical structures and processes in the region \cite{kivelson2004magnetospheric, jia2010magnetic, harris2021multi}.

%---------------------------------------------------------------------------------------
\subsection{Parameter Importance}
\begin{figure}[h]
    \centering
    \includegraphics[width=0.6\textwidth]{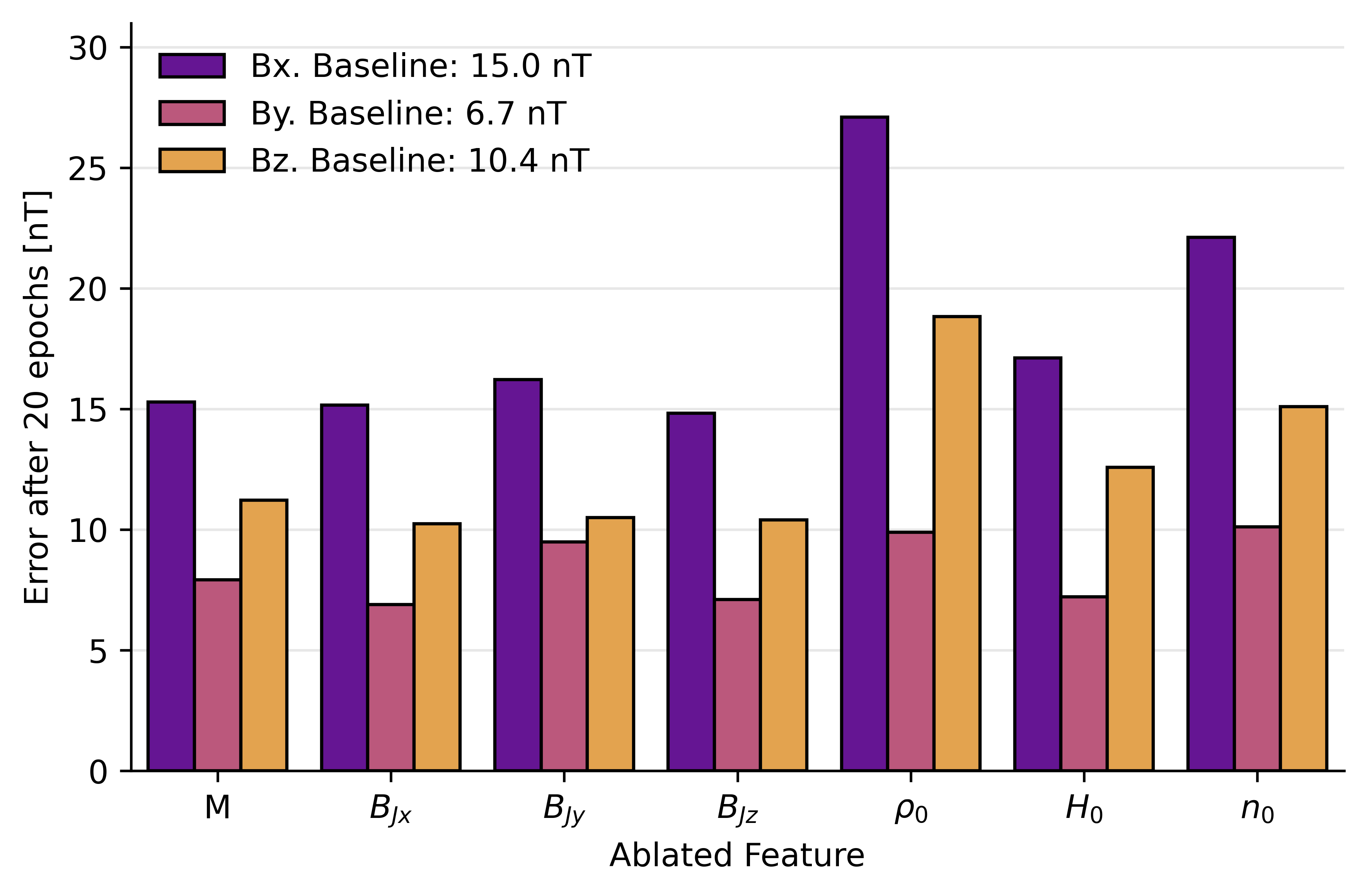}
    \caption{\textbf{Estimating feature importance.} Ablation of the nine physical parameters showing their impact on MAE for $B_x$, $B_y$, and $B_z$.}
    \label{fig9}
\end{figure}

Finally, we use LEAP to identify which physical parameters have the greatest influence on the MHD outputs. This analysis is performed to strategically enhance targeted future MHD campaigns in under-sampled regions of the parameter space. Feature importance is assessed by systematically ablating each physical parameter one at a time and measuring the resulting MAE. Geometric inputs $(x, y, z, R)$ are excluded because they are not adjustable in the MHD model. We also fix the runs to 20 epochs to streamline the analysis and reduce compute cost. Figure~\ref{fig9} shows that removing $\rho_0$ has the largest effect in $B_x$, increasing the error from 15.0 nT $\rightarrow$ 27.1 nT (+80.7\%). Next are $n_0$ and $H_0$ whose removal increases the $B_x$ error by +47.5\% and +14.2\%, respectively. Among the features, the error pattern remains consistent, with the ablated features mainly affecting $B_x$, then $B_z$ and lastly $B_y$. An earlier sensitivity analysis using the parent MHD code \cite{harris2022multi} also identified $\rho_0$ and $H_0$ as primary drivers of the magnetic response in $B_z$, which our results broadly corroborate. This rapid, systematic parameter ranking is a core proposition of an ML-MHD loop, allowing us to prioritize expensive MHD runs in parameter space where the model's sensitivity is highest.

%---------------------------------------------------------------------------------------
%---------------------------------------------------------------------------------------
%---------------------------------------------------------------------------------------
%----------------------------------  D I S C U S S I O N  ------------------------------
%---------------------------------------------------------------------------------------
%---------------------------------------------------------------------------------------
%---------------------------------------------------------------------------------------
%1375 words
\section{Discussion}\label{sec-discuss}
%Recap
In this study, we introduce LEAP: a transformer-based surrogate for quick and accurate predictions of the magnetic field in the vicinity of Europa. On the test set, it reproduces all magnetic field components within an error of $\pm$2.6 nT. For the Galileo $E4$ and $E14$ flybys of Europa, LEAP is comparable in performance to the existing state-of-the-art MHD model code. Crucially, the emulator runs near-instantly, enabling novel analyses such as parameter surveys and probabilistic parameter estimations via Bayesian inference. LEAP can also help inform future MHD runs by identifying the most important parameters. We now discuss some of these results as well as the current limitations of the framework.

%Existing works
The emulation of MHD codes with ML in solar system science is still an emerging field with few examples in the domain. To our knowledge, \cite{kataoka2024machine} was the first to successfully do so with the SMRAI model of the REPPU MHD code \cite{tanaka1994finite}. More recently, \cite{holmberg2025graph} created a surrogate of the hybrid Vlasiator code which treats electrons as a fluid and ions as kinetic \cite{von2014vlasiator}. In both instances, they chose time-series specific architectures: an echo-state network and graph neural network, respectively. In planetary science there is neither the resource nor demand for such frequent runs, and so other architectures have been explored. \cite{azari2023simulation} illustrated a Bayesian surrogate of the Duling MHD code of the Ganymede system \cite{duling2022ganymede}, but these results are more preliminary by comparison. A direct quantitative comparison is not feasible here due to differences in problem formulation and target variables, but we hope that such technologies will continue to be explored and expanded further by the community.

%Performance on $E4$ and $E14$
As shown in Figure \ref{fig5}, speed accompanies performance: LEAP generalizes to unseen geometry configurations, often matching the MHD baseline. A question that naturally arises is: could LEAP solely be trained on observational data, as this is the benchmark we are validating against? While the scarcity of observational data currently precludes this, a future multi-modal approach incorporating both MHD and magnetometry data is a natural next step. The sequence-aware nature of the Transformer is well suited to integrate the heterogeneous data \cite{huang2020tabtransformer} from in-situ magnetometers and simulated values from MHD. That said, such a marriage is not without challenges. Observational data may have to be down-scaled to match the MHD data, potentially erasing useful fine-scale features (see `raw' in Fig \ref{fig5}). Furthermore, adding observational data contaminates the validation process and would bias the feature-importance analysis (Fig. \ref{fig9}) needed for targeted future MHD runs. 

%Inverse Modeling
The Bayesian inversion method (Fig.~\ref{fig6}) demonstrates that simulated physical parameters can be recovered via a transformer surrogate. Our results also highlight the need for such an inversion, since the initial MHD estimates for future MHD runs can be improved. Currently, the Bayesian method estimates MHD \textit{simulation} parameters, but it could be extended to \textit{in-situ} observations, enabling real-time estimations of plasma conditions from ECM \cite{kivelson2023ecm} and PIMS \cite{westlake2023plasma}. A direct inversion of observations was not performed here because a full validation chain would require taking the maximum posterior parameters inferred from the surrogate and re-running a forward MHD simulation \cite[e.g.,][]{harris2021multi} to verify consistency with the observations. This end-to-end validation step is not possible at present and therefore remains future work. Such an approach would provide true probabilistic constraints on the plasma parameters in an environment poorly constrained by observations. This is crucial because (1) the plasma is highly inhomogeneous and changes on the order of minutes \cite{saur2010induced}, (2) the plasma contribution ($B_{plasma}$ in Eq. \ref{eq:B_obs}) can affect the estimation of Europa's ice shell thickness \cite{biersteker2023revealing}, which forms a key part of Europa Clipper's science objectives, and (3) because the plasma environment affects the excitation moments applied to the moon, thus confounding efforts to isolate the induced field ($B_{ocean}$ in Eq. \ref{eq:B_obs}) \cite{styczinski2024planetmag}. Finally, LEAP contributes to a growing set of inversion methods being developed for Europa \cite{biersteker2023revealing, winkenstern2025inversion, styczinski2023planetprofile}.

%Forward Modeling 
The parameter survey applied to the JUICE flybys of Europa (Fig.~\ref{fig7}) shows that LEAP has sufficient fidelity to capture known physical phenomena, such as magnetic pile-up and stretching, and Alfvén wing structures. This is further corroborated by predictions of the magnetic field along the flybys in the test set (Fig \ref{fig2}). 
The local plasma density, which is poorly constrained observationally, can be determined from the Alfvén wing angle \cite{neubauer1980nonlinear}. As LEAP runs rapidly, this presents a novel opportunity to perform a statistical analysis of plasma density in the downstream region of Europa, using the radio occulations measurements as an upper bound \cite{kliore1997ionosphere}. That said, it is not known if LEAP can reproduce the entire volume (akin to MHD) from only the flyby trajectories, so further investigations may be required. Finally, LEAP now enables an investigation of PIFs for various interior structures of Europa ahead of Clipper's arrival in 2031, when jointly used with other statistical methods like those recently developed for Callisto \cite{cochrane2025stronger}. Such an investigation could allow for a greater understanding of dominant plasma interactions (useful for PIMS \cite{westlake2023plasma}) and could further aid the magnetic induction experiment \cite{kivelson2023ecm}.

%Parameter importance
Figure~\ref{fig6} summarizes the feature importance revealed by the ablation test. The goals here were twofold: (1) to benchmark against and expand earlier sensitivity work \cite{harris2022multi} by adding $B_x$ and $B_y$, and four additional parameters; and (2) to identify parameters that should be prioritized in future MHD campaigns. To strengthen LEAP's generalization and to optimize future MHD campaigns, we recommend focusing on new runs that vary (i) magnetospheric O$^+$ mass density ($\rho_0$), (ii) atmospheric scale height ($H_0$), (iii) atmospheric surface density ($n_0$) as these parameters produce the largest degradation when removed. Although the magnetic moment ($M$) and background magnetic field values ($B_{J, xyz}$) appear less influential, their importance is expected to increase with new observational data acquired by Clipper \cite{westlake2023plasma, kivelson2023ecm} and JUICE \cite{masters2025magnetosphere}, avoiding a future redesign of the model. Lastly, the above highlights a mutually beneficial relationship: LEAP narrows the MHD parameter space to the most impactful regions, enabling more efficient and targeted simulations, while additional high-quality MHD outputs further improve future versions of LEAP. This can then be repeated with future ablation studies, and the cycle continues, becoming progressively more useful.

%limitations remain
LEAP, like all surrogates, inherits the assumptions and omissions of its parent model. The BATSRUS Europa MHD model \cite{harris2021multi, harris2022multi} performs very well compared to the \textit{E4} and \textit{E14} observations (Figure \ref{fig5}), but this might not extend to all past and future flybys. For example, Juno's 2022 flyby revealed electron beams associated with Europa's leading hemisphere \cite{allegrini2024electron}, which have since been incorporated into a single-fluid MHD code \cite{cervantes2025mhd}. That work showed improved fits to some components of the magnetic field when the electron beams were included in the model, though it remains unclear if this beam-inclusive model would improve the observation-simulation alignment on other flybys, such as \textit{E4} and \textit{E14}. Similarly, diurnal variability in Europa's O$_2$ exosphere -- known as the `dawn-over-dusk asymmetry' \cite{oza2019dusk} -- is represented in hybrid codes such as AIKEF \cite{addison2024magnetic}, but not in the BATSRUS MHD model used in this work. These omissions may become important for future comparisons, but they do not appear to be critical for the current benchmarking. The BATSRUS MHD model \cite{harris2021multi} remains the only model to simulate three ion fluids, and its strong agreement with the Galileo in-situ data suggests multiple fluids are a driving factor for an accurate fit (See Fig. \ref{fig5}). Also, LEAP is currently restricted in scope. Unlike full fluid or hybrid models \cite{saur1998interaction, schilling2008influence, jia2018evidence, harris2022multi, addison2024magnetic, cervantes2025mhd}, which output a suite of useful plasma quantities (e.g., density, pressure, velocity, etc), the current version of LEAP only predicts magnetic fields. While magnetic fields are of primary relevance for the induction experiment \cite{kivelson2023ecm} and ocean/ice characterization \cite{vance2023investigating}, our flexible transformer framework (Fig. \ref{fig1}, Eqs. \ref{eq:flyby-traj}-\ref{eq:regression_head}) makes it straightforward to train companion models that simultaneously predict a range of plasma parameters. Such models could serve as inputs for downstream analyses. For example, MHD simulations are often used to drive test-particle calculations; thus, LEAP-like frameworks could provide broader insight into the behavior of energetic particles \cite{meitzler2023investigating} and their role in surface modification processes \cite{nordheim2022magnetospheric}.

%Final bit on survey, estimation, importance.
Lastly, Europa Clipper and JUICE are set to arrive at Europa from 2031 \cite{Grasset2013, pappalardo2024science}; their observations may provide a way to validate the parameter estimation and survey, as well as to settle the questions regarding the most important plasma parameters in the observed magnetic field.

%---------------------------------------------------------------------------------------
%---------------------------------------------------------------------------------------
%---------------------------------------------------------------------------------------
%---------------------------------------------------------------------------------------
%---------------------------------------------------------------------------------------
%---------------------------------------------------------------------------------------
%---------------------------------------------------------------------------------------
%-------------------------------------------------------------------------------------
%Conclusion, but not a conclusion section (2.0)
In this study, we have demonstrated how transformers can act as rapid and accurate surrogates for multi-fluid magnetohydrodynamic models. This acceleration enables parameter estimation via simulated based inference, providing the potential for real-time and probabilistic estimates of plasma conditions at Europa, representing a shift in the analysis of moon–magnetosphere environments. This capability is crucial for characterizing plasma interactions which confound our ability to magnetically sound ocean worlds like Europa. The fidelity of LEAP is validated by its ability to reproduce established Galileo flybys, as well as key physical phenomena, such as Alfvén wing draping and magnetic pile-up. We also propose an adaptive loop whereby MHD and ML models iteratively refine one another by exploring under-sampled parameter spaces. We anticipate that this framework could aid mission planning and be applied to other bodies of interest, such as Neptune or Uranus \cite{frazier2020trident, girija2023flagship}, whose plasma environments remain poorly constrained. These results establish a scalable framework for ML-accelerated space plasma modeling.

%---------------------------------------------------------------------------------------
%---------------------------------------------------------------------------------------
%---------------------------------------------------------------------------------------
%----------------------------------  M E T H O D S  ------------------------------
%---------------------------------------------------------------------------------------
%---------------------------------------------------------------------------------------
%---------------------------------------------------------------------------------------
%---------------------------------------------------------------------------------------

%No word limit
%TC:ignore
\section{Methods}\label{sec-method}
The next section describes the MHD training data and underlying physical processes, as well as the construction of LEAP—from extracting and rotating flyby trajectories within the MHD simulations, to transformer training protocols and parameter estimation via Bayesian inference and Markov-chain Monte Carlo. The overall workflow is summarized in Figure~\ref{fig1}.

%---------------------------------------------------------------------------------------
\subsection{Magnetohydrodynamic Training Data}
LEAP is trained exclusively on outputs of a three-dimensional multi-fluid magnetohydrodynamic (MHD) code \cite{harris2021multi, harris2022multi, nordheim2022magnetospheric}, based on the BATSRUS framework \cite{glocer2009multifluid, toth2012adaptive, gombosi2021swmf}. This code self-consistently solves for plasma and magnetic field properties in the vicinity of Europa within the Jovian magnetospheric environment. The simulation includes four coupled fluids corresponding to electrons, magnetospheric $\mathrm{O}^+$ ions, and ionospheric/exospheric $\mathrm{O}^+$ ions and $\mathrm{O}_2^+$ ions \cite{harris2021multi, harris2022multi}. The ambient plasma flow originating from the Io plasma torus streams onto Europa from the upstream simulation boundary \cite{kivelson2004magnetospheric}. Ion populations associated with Europa’s atmosphere are generated via electron impact ionization, photoionization, and charge exchange processes \cite{kliore1997ionosphere, plainaki2013exospheric}.

The numerical formulation of the MHD model is described in detail by \cite{rubin2015self}, with supplementary details in \cite{harris2021multi, harris2022multi}. In its native format, the 3D MHD domain is on a spherical mesh logarithmically spaced in the radial direction, with grid resolutions ranging from 15 km at Europa's surface (0.01 Europa radii, $\mathrm{R_E}$) to 15,700 km (10 $\mathrm{R_E}$) at the simulation outer boundary which is $\sim$ 28 $\mathrm{R_E}$ away from the moon. For consistency and compatibility with the machine learning framework, the MHD output is interpolated onto a uniform Cartesian grid with 51 km spacing in all three spatial dimensions. Each data cube spans $x,z \in [-5,5]\,\mathrm{R_E}$ and $y \in [-2.5,2.5]\,\mathrm{R_E}$, yielding approximately 300x150x300 = 13.5 million grid cells per snapshot.
The MHD simulations provide three-dimensional magnetic field components $(\mathbf{B_x}, \mathbf{B_y}, \mathbf{B_z})$. In this study, the background Jovian magnetic field is removed from these outputs, and the resulting perturbations $(\Delta B_x, \Delta B_y, \Delta B_z)$ serve as the target values for the LEAP emulator.

The MHD training data were selected to represent two categories: (1) real flybys for which observational data constrain the upstream and boundary conditions, and (2) simulated flybys designed to span the expected variability arising from Jupiter’s 11.2-hour synodic rotation period and Europa’s 85.2-hour orbital period \cite{kivelson2004magnetospheric, jia2010magnetic}. The latter category is illustrated in the third panel of Figure~\ref{figx1}, which demonstrates the wide range of current sheet configurations expected during Europa Clipper and JUICE flybys. Together, these datasets cover the anticipated extrema of the relevant parameter space based on prior Galileo and Juno observations and established MHD–observation fits \cite{harris2021multi, harris2022multi}. All training simulations assume either no ocean induction (0\% efficiency) or a 100\% efficient induction consistent with a conductive subsurface ocean. The underlying MHD training data have been previously analyzed in \cite{harris2021multi, harris2022multi, nordheim2022magnetospheric}.

\subsection{Induction at Europa}
Europa resides within Jupiter’s rapidly rotating magnetosphere, where time-varying plasma flows and magnetic fields interact with the moon’s conductive interior and surrounding plasma environment. Magnetic field perturbations arise from two physically coupled processes: (1) plasma interaction effects driven by the deflection and deceleration of magnetospheric flow around Europa, and (2) electromagnetic induction within Europa’s subsurface saline ocean \cite{khurana1998induced}. Because these processes are coupled, spacecraft magnetometer observations cannot disentangle individual contributions and instead measure their combined effect.
Magnetic field perturbations associated with the plasma interaction arise from  \cite{saur2010induced},

\begin{equation}
    \frac{\partial \mathbf{B}}{\partial t} = \nabla \times (\mathbf{v} \times \mathbf{B}) - \nabla \times \left( \eta \, \nabla \times \mathbf{B} \right),
    \label{eq:B-convection}
    \end{equation}
    
where $\mathbf{v}$ is the plasma flow velocity and $\eta = 1/(\mu_0 \sigma)$ is the magnetic diffusivity. The first term represents convective magnetic field perturbations driven by plasma flow changes around Europa with respect to the background flow, while the second term represents diffusive processes associated with finite electrical conductivity \cite{saur2010induced, styczinski2024planetmag}. Within Europa’s conductive subsurface ocean, magnetic field evolution is dominated by diffusion and can be approximated by \cite{saur2010induced, khurana2009electromagnetic}

\begin{equation}
    \frac{\partial \mathbf{B}}{\partial t} \approx \eta \nabla^2 \mathbf{B},
    \label{eq:B-diffusion}
\end{equation}
Closest to an inducting body, the magnetic response is strongest, decaying as a function of $1/r^3$. The magnetic field measured by a spacecraft near Europa can be expressed schematically as \cite{schilling2008influence, saur2010induced, cochrane2024detecting}

\begin{equation}
    \mathbf{B}_{\mathrm{obs}} =
    \mathbf{B}_J +
    \mathbf{B}_{\mathrm{plasma}} +
    \mathbf{B}_{\mathrm{ocean}}(\mathbf{B}_J) +
    \mathbf{B}_{\mathrm{ocean}}(\mathbf{B}_{\mathrm{plasma}}) +
    \mathbf{B}_{\mathrm{sc}} + \epsilon ,
    \label{eq:B_obs}
\end{equation}
\newline

where $\mathbf{B}_J$ is the background Jovian magnetic field, $\mathbf{B}_{\mathrm{plasma}}$ represents perturbations arising from plasma interaction processes ($\approx$ Eq. \ref{eq:B-convection}), and $\mathbf{B}_{\mathrm{ocean}}$ denotes magnetic fields induced within Europa’s conductive ocean in response to external time-varying fields ($\approx$ Eq. \ref{eq:B-diffusion}). The term $\mathbf{B}_{\mathrm{sc}}$ accounts for spacecraft-generated magnetic disturbances \cite{cochrane2023magnetic}, and $\epsilon$ represents residual measurement uncertainties relating to pointing, timing, calibration, and trajectory errors \cite{cochrane2024detecting}.

Equation~\ref{eq:B_obs} highlights two key points relevant to this study. First, plasma interaction and ocean induction are inherently coupled processes and must be treated self-consistently to comprehensively describe the magnetic field \cite{saur2010induced}. Second, accurate characterization of Europa’s subsurface ocean properties, including conductivity and salinity \cite{vance2023investigating, styczinski2023planetprofile}, requires robust modeling of the plasma interaction environment.

In its native form, the MHD output includes the full magnetic field contributions shown in Equation~\ref{eq:B_obs}. Because the present study focuses on magnetic perturbations associated with plasma interaction and induction, we subtract the background Jovian magnetic field from each simulation snapshot. Accordingly, the machine learning emulator is trained to predict magnetic field perturbations
\begin{equation}
\Delta \mathbf{B} = \mathbf{B}_{\mathrm{MHD}} - \mathbf{B}_J .
\end{equation}
LEAP therefore only learns the plasma-driven and induced magnetic field perturbations. To produce physically realistic magnetic field predictions along spacecraft trajectories, the background Jovian magnetic field is added back using the Khurana field model \cite{khurana1997euler}. This background field model will be upgraded as Europa Clipper and JUICE magnetometer data become available \cite{kivelson2023ecm}.

In summary, the outputs of LEAP provide rapid predictions of $\Delta \mathbf{B}$. Future studies may use these outputs to better constrain Europa’s interior structure, including ocean conductivity and layering, through joint plasma–induction analyses \cite{styczinski2023planetprofile}.

%---------------------------------------------------------------------------------------
\subsection{Extracting and Padding Flybys}
%Extracting Flybys
Instead of learning the full 13.5 million data points within the entire MHD domain, we carve out the flybys of past and future missions at Europa (Galileo, Juno, Europa Clipper, and JUICE) from the data cube (See Fig. \ref{figx1}). The reason for this is twofold. We are mostly interested in the expected magnetic field conditions that JUICE and Clipper may experience, rather than the entire physics of the system. Secondly, by only learning the flybys, we reduce the dataset from 13.5 million values to approximately 60,000 -- a compression factor of about 225$\times$ -- without compromising on accuracy and significantly reducing the computational load. The tour designs are well established \cite{boutonnet2024designing, cangahuala2025europa} and the ephemerides are publicly available via the \texttt{NAIF SPICE} kernels \footnote{https://naif.jpl.nasa.gov/pub/naif/}. We then map the flyby ephemerides to the same 51 km voxel grid to match the MHD data. Each flyby is defined as an ordered sequence of discrete time-aligned points: 

\begin{equation}
     %\tau_i = \{{(x_t, y_t)}\}_{t=1}^{T_i}, 
     \tau_i = \{(x_t, y_t)\}_{t=1}^{T_i},
     \label{eq:flyby-traj}
\end{equation}
\newline

where the input vector $x_t$ contains the voxelized spacecraft positions $x, y, z$; the position vector magnitude $R$; and physical parameters ($\boldsymbol{\theta}$) shown in Figure \ref{fig9}. The target vector $y_t$ contains the magnetic field components ($B_x, B_y, B_z$) which we want to predict. We also predict the total magnitude of the magnetic field, which is not formally learned, but derived from $B = \sqrt{(B_x^2 + B_y^2 + B_z^2)}$. $T$ is the trajectory length for flyby $i$. There are a total of 63 flybys of Europa across the four missions: Galileo (11; 1996–2003), Juno (1; 2022), Europa Clipper (49; 2031–2033), and JUICE (2; 2031). We augment the data by applying the 24 orientation symmetries of a cube to each $\tau_i$ (see Figs \ref{figx2}-\ref{figx3}). This physics-informed data augmentation strategy leverages the geometrical invariance of the MHD governing equations to synthesize a robust training set. Each rotation is represented by a 3 $\times$ 3 permutation matrix, e.g.

\begin{equation}
R_1 =
\begin{bmatrix}
1 & 0 & 0 \\
0 & 0 & -1 \\
0 & 1 & 0
\end{bmatrix},
\quad
R_2 =
\begin{bmatrix}
0 & 0 & 1 \\
0 & 1 & 0 \\
-1 & 0 & 0
\end{bmatrix},
\label{eq:cube_rot_examples}
\end{equation}
\newline

which respectively map $(x, y, z) \mapsto (x, -z, y)$ and $(x, y, z) \mapsto (z, y, -x)$. Thus we obtain 24 additional versions of $\tau$, including the B field. Applying all 24 cube rotations yields a total of $32 \text{ volumes} \times 63 \text{ flybys} \times 24 \text{ rotations } = 48{,}348 \text{ maximum sequence samples}.$ The total number of learnable time-steps across all samples is approximately 9 million. While modest, this dataset is expected to grow over the next decade during the cruise and operational phases of Europa Clipper and JUICE. This study demonstrates that even a relatively limited training set can generate physically grounded predictions and, by design, the LEAP framework scales easily with increasing data availability.

%Need to discuss the sampling. What are the 32? These are not just random, but designed to cover the entire expected rnage for the missions. E.g. we book-end the paramter space. To be clear, the limtied samples means we cannot predic tbeyond this rnage. If we experience this, new MHD will need to be trained.

%Flyby Padding
Each $\tau_i$ has variable length—ranging from $T$ = 89 (Juno, PJ45) to $T$ = 557 (JUICE, 8E2). To ensure consistent tensor lengths, we pad all trajectories to the maximum observed length within a batch, denoted by $T_{\max}$. Each flyby is represented by three tensors:

\[
\mathbf{X} \in \mathbb{R}^{B \times T_{\max} \times d_{\text{in}}}, \quad
\mathbf{Y} \in \mathbb{R}^{B \times T_{\max} \times d_{\text{out}}}, \quad
\mathbf{M} \in \{0,1\}^{B \times T_{\max}}.
\]
\newline

where $\mathbf{X}^{(i)}$ is the padded input sequence of length $T_{\max}$ with feature size $d_{\text{in}}$, $\mathbf{Y}^{(i)}$ the corresponding padded target sequence with feature size $d_{\text{out}}$, $\mathbf{M}^{(i)}$ a binary mask indicating valid positions ($1$) versus padding ($0$), and $B$ is the batch size (here, $B=8$). This representation preserves temporal ordering along each flyby trajectory, enables batching across missions, and allows sequence models (e.g., Transformers) to ignore padding through the mask $\mathbf{M}$. 

%---------------------------------------------------------------------------------------
\subsection{Transformer}
We adopt a lightweight encoder-only Transformer \cite{vaswani2017attention} to map input flyby sequences to predicted magnetic field outputs. 
%others
The echo state network and graph neural network approaches used in the space weather community \cite{kataoka2024machine, holmberg2025graph} were not applicable here, as our model operates on steady-state representations rather than continuous time-series inputs. We also tested a vanilla multi-layer perceptron (MLP), but it produced noisy predictions with higher errors. This is expected because MLPs treat each point independently and cannot examine the entire flyby trajectory. Recurrent models (e.g., LSTMs) were also considered but not adopted due to their sensitivity to sequence length (which varies from $T=89$ to $T=557$) and vanishing-gradient behavior. Transformers are better suited to this problem because they process the entire trajectory simultaneously using self-attention. The ability of self-attention to establish dependencies between all time-steps $t$ in the sequence is critical for accurately modeling large-scale, non-local phenomena like Alfvén wings \cite{neubauer1980nonlinear, volwerk2007europa} and the global magnetic pile-up geometry \cite{harris2021multi}, which are determined by the plasma interactions across the entire flyby domain. We choose an encoder-only design because the task involves predicting the magnetic field along the observed trajectory, where input and output sequences are aligned, making an encoder–decoder architecture unnecessary. %Encoder–decoder architectures are not required here, as we are not generating a future or shifted sequence, and the encoder alone is sufficient to capture the long-range dependencies needed for accurate regression. 
We omit sinusoidal or learnable positional encoding, since the spacecraft positions ($x,y,z,R$) are explicitly included in the input vector, providing positional and geometric information. The input vector at each trajectory step, $\mathbf{X}^{(i)} \in \mathbb{R}^{T_{\max} \times d_{\text{in}}}$, is linearly projected to $d_{\text{model}}$, which is the dimensionality of the embedding space \cite{vaswani2017attention}:

  \begin{equation}
    \mathbf{h}_t^{(0)} = \mathbf{W}_{\text{proj}} \mathbf{x}_t + \mathbf{b}_{\text{proj}}, 
    \label{eq:input_proj}
  \end{equation}
\newline

%d_model the dimensionality of the embedding space and model representations)
where $\mathbf{W}_{\text{proj}} \in \mathbb{R}^{d_{\text{model}} \times d_{\text{in}}}$ and $\mathbf{b}_{\text{proj}} \in \mathbb{R}^{d_{\text{model}}}$. The projected trajectories $ \mathbf{H}^{(0)} =\{\mathbf{h}_t^{(0)}\}_{t=1}^T$ are passed through a stack of $L$ Transformer encoder layers. Each layer contains $H$ heads and a vanilla MLP feedforward network. The final encoder output is

\begin{equation}
    \mathbf{H}^{(L)} = \text{TransformerEncoder}\left( \mathbf{H}^{(0)} \right),
    \label{eq:encoder}
\end{equation}
\newline

where $\mathbf{H}^{(L)} \in \mathbb{R}^{T \times d_{\text{model}}}$. Finally, a linear function maps each hidden state at timestep $t$ to the output space. Since LEAP predicts all 3 magnetic field components, the output layer is vector valued 

\begin{equation}
    \hat{\mathbf{y}}_t = \mathbf{W}_{\text{out}} \mathbf{h}_t^{(L)} + \mathbf{b}_{\text{out}}, 
    \label{eq:regression_head}
\end{equation}
\newline

where $\hat{\mathbf{y}}_t \in \mathbb{R}^3$, $\mathbf{W}_{\text{out}} \in \mathbb{R}^{3 \times d_{\text{model}}}$, and $\mathbf{b}_{\text{out}} \in \mathbb{R}^3$. In this study, we deliberately employ a lightweight architecture and set $H = 2$, $L=2$, $d_{\text{model}} = 32$, and the inner-layer dimensionality of the feed-forward network, $d_\text{ff}$, to 4096. This deviates from the original convention whereby $d_\text{ff} = d_{\text{model}} \times 4$ \cite{vaswani2017attention}. We also experimented with $d_{\text{model}} \in \{64, 128, 512\}$, $L \in \{3, 4\}$, $H \in \{3, 4\}$, $d_\text{ff} \in \{128, 256, 1024\}$, but observed declining performance, confirming that minimal complexity was the optimal solution for this domain.

%--------------------------------------------------------------------------------------
\subsection{Training Objective}
We design a physics-constrained loss function that captures the radial dependence implied by Equations~\ref{eq:B-convection} and \ref{eq:B-diffusion}, the anisotropy of magnetic perturbations parallel and perpendicular to the background Jovian field, and ensures a smooth spatial variation of magnetic fields along a spacecraft trajectory. %The latter is not intended to enforce the full Maxwell or MHD equations, but to regularize the predictions toward physically plausible behavior \cite{holmberg2025graph}.

Let $\widehat{\mathbf{B}}_i$ and $\mathbf{B}_i$ denote the predicted and true magnetic perturbation vectors (i.e., $\Delta \mathbf{B}$) at sample index $i$, and let $\mathbf{B}_{J,i}$ denote the background Jovian magnetic field at the same location. The perturbation error is
\begin{equation}
\Delta \mathbf{B}_i = \widehat{\mathbf{B}}_i - \mathbf{B}_i .
\end{equation}

To capture anisotropic error structure, we decompose $\Delta \mathbf{B}_i$ into components parallel and perpendicular to the local background field direction. We define the unit vector as
\begin{equation}
    \hat{\mathbf{b}}_i = \frac{\mathbf{B}_{J,i}}{\left\lVert \mathbf{B}_{J,i} \right\rVert + \varepsilon},
\end{equation}
where $\varepsilon$ is a numerical stability constant set to 10$^{-8}$ to avoid a potential division by zero. The parallel and perpendicular components are
\begin{align}
    \Delta \mathbf{B}_{\parallel,i} &= \left(\Delta \mathbf{B}_i \cdot \hat{\mathbf{b}}_i\right)\hat{\mathbf{b}}_i, \\
    \Delta \mathbf{B}_{\perp,i} &= \Delta \mathbf{B}_i - \Delta \mathbf{B}_{\parallel,i}.
\end{align}
We quantify their magnitudes using an $\ell_1$ norm,
\begin{equation}
    e_{\parallel,i} = \left\lVert \Delta \mathbf{B}_{\parallel,i} \right\rVert_{1},
    \qquad
    e_{\perp,i} = \left\lVert \Delta \mathbf{B}_{\perp,i} \right\rVert_{1}.
\end{equation}

To account for the reduced magnetic variability far from Europa, we apply a bounded radial weighting. Let $R_i$ denote the spacecraft distance from Europa and let $R_{\max}$ be the maximum distance in the trajectory window. We define
\begin{equation}
    w_i = \mathrm{clip}\!\left(
    1 - \frac{R_i}{R_{\max} + \varepsilon},\,0,\,1\right).
    \label{eq:training-obj-radial}
\end{equation}

where $\mathrm{clip}(\cdot)$ constrains values to $[0,1]$ to keep the loss response steady. The error is then computed as 
\begin{equation}
    \mathcal{L}_{\mathrm{\parallel, \perp}} =
    \frac{1}{N}\sum_{i=1}^{N} w_i \left(\alpha \, e_{\parallel,i} + \beta \, e_{\perp,i}\right),
    \label{eq:training-obj-par-perp}
\end{equation}

where $\alpha$ and $\beta$ control the relative weighting of the parallel and perpendicular error components. In this study, we set $\alpha = 0.5$ and $\beta = 1.0$, placing greater emphasis on transverse perturbations. In addition, we introduce a trajectory-based regularization term motivated by the constraint $\nabla \cdot \mathbf{B} = 0$ \cite{holmberg2025graph} that penalizes rapid point-to-point variations in the predicted magnetic perturbations, which are non-physical or represent noise in the MHD domain.

\begin{equation}
\mathcal{L}_{\mathrm{grad}} =
\frac{1}{N-1}\sum_{i=1}^{N-1}
\left|
\sum_{k \in \{x,y,z\}}
\frac{\Delta \widehat{B}_{k,i}}{\left\lVert \Delta \mathbf{r}_i \right\rVert + \varepsilon}
\right|.
\end{equation}

where $\mathbf{r}_i$ denotes the spacecraft position, $\Delta \mathbf{r}_i = \mathbf{r}_{i+1} - \mathbf{r}_i$, and $\Delta \widehat{\mathbf{B}}_i = \widehat{\mathbf{B}}_{i+1} - \widehat{\mathbf{B}}_i$.

The full training objective is then given by the sum
\begin{equation}
    \mathcal{L}_{\mathrm{tot}} =
    \mathcal{L}_{\mathrm{\parallel, \perp}} + \lambda_{\mathrm{grad}} \, \mathcal{L}_{\mathrm{grad}},
    \label{eq:training-obj-combined}
\end{equation}
where $\lambda_{\mathrm{grad}}$ controls the strength of the trajectory-based regularization and is set to $\lambda_{\mathrm{grad}} = 0.1$. This choice reflects the fact that spatial derivatives perpendicular to the trajectory are not available to the emulator, enforcing this constraint too strongly would be nonphysical. Instead, the regularization term acts primarily to suppress non-physical high-frequency fluctuations in the predicted perturbations while weakly encouraging behavior consistent with the divergence-free nature of magnetic fields assumed by the underlying MHD model \cite{schilling2008influence, harris2021multi}.
Lastly, we adopt a vanilla mean absolute error (MAE) for assessing the test set performance (Fig. \ref{fig3}a), and for quantifying the LEAP predictions and the MHD simulations with respect to the in-situ Galileo observations (Fig. \ref{fig5}).

\subsection{Training Protocols}
%train-test-split
We randomly split the trajectories into 80\%--10\%--10\% train/evaluate/test sets, fixing the seed to 42 for reproducibility. We also experimented with 70-20-10 but performance on the test-set deteriorated. This is somewhat expected as our dataset is limited ($\sim$ 48k trajectories) and this should improve as much more MHD are generated over the cruise and operational phases (10+ years). We perform 5-fold cross-validation over unique trajectory IDs ($\tau$) using \texttt{KFold} (\texttt{n\_splits}=5, \texttt{shuffle=True}, \texttt{random\_state}=42) to quantify generalization across trajectories; the resulting fold-wise errors are stable, with $\mathrm{MAE}=1.76\ \mathrm{nT}\pm0.17\ (9.6\%)$. We also check to ensure there are real flybys within the validation and test sets, not just rotations (Eq. \ref{eq:cube_rot_examples}). The Galileo $E4$ and $E14$ flybys from the Harris study \cite{harris2021multi} are held out in their native format (RO4) for dedicated testing and validation (see Figure \ref{fig5}). 

%Adding Noise
%We add noise to the position inputs to represent uncertainty per Cochrane. We had rouhglt 15 km by randomly shifting the components for every batch

%Leakage
Next, we introduce an additional cross-validation schema to assess potential spatial data leakage. Firstly, the rotation of the flybys as part of the data augmentation process does not return any identical trajectories as this is forbidden by the permutation matrix (See Eq. \ref{eq:cube_rot_examples}). We illustrate this for the $E4$ and $E14$ flybys (Figs. \ref{figx2}-\ref{figx3}), but the same logic applies to all the flybys (Fig. \ref{figx1}). Secondly, we verify that no duplicate coordinates $(x, y, z) \in \mathbb{R}^3$ occur (0\%) within any $\tau$ -- which would indicate that trajectories overlap. Finally, we quantify any spatial information leakage arising from the `neighbors of neighboring cells' that \textit{might} share target-value information. We use a \texttt{KDTree} nearest-neighbor search to identify all points with neighbors within a distance of two bins or 102 km ($r$):

\begin{equation}
    \left\lVert \mathbf{q}_j - \mathbf{p}_i\right\rVert < r,
\end{equation}
\newline

where $\mathbf{p}_i$ denotes a test point, and $\mathbf{q}_j$ is its neighboring points. We exclude any neighbors that belong to the same trajectory and discard cases with fewer than five common neighboring points within $r$. This criterion ensures that we only consider trajectories that run approximately in parallel, rather than those that simply intersect at a few points (i.e. crisscross). The average trajectory length, $\langle T \rangle$, is 191, so 5 points represents $<$ 3\% of the overall trajectory, $\tau_i$. 

Formally, let $\mathcal{K}_j$ denote the set of spatial neighbors $\mathbf{q}_j$ on different trajectories that have at least five common points with $\mathbf{p}_i$ within $r$. We define $\mathcal{O}(\mathbf{p}_i, \mathbf{q}_j)$ as an indicator function for such pairs:

\[
\mathcal{O}(\mathbf{p}_i, \mathbf{q}_j) =
\begin{cases}
1 & \text{if trajectories satisfy the spatial-neighbor criteria,}\\
0 & \text{otherwise.}
\end{cases}
\]

The spatial soft leakage fraction is then defined as

\begin{equation}
    f_{\mathrm{soft}} = \frac{1}{N_{\mathrm{test}}} \sum_{j=1}^{N_{\mathrm{test}}} \mathbb{I}\!\left(\sum_{i \in \mathcal{K}_j} \mathcal{O}(\mathbf{p}_i, \mathbf{q}_j) > 0 \right),
    \label{eq:spatial-leakage}
\end{equation}
\newline

where $N_{\mathrm{test}}$ is the total number of test points across every $\tau_i$. In short, $f_{\mathrm{soft}}$ measures the fraction of test points that could plausibly have been indirectly `seen' during training due to spatial proximity of neighboring trajectories with potentially similar values. Our analysis yields $f_{\mathrm{soft}} \approx 3.9\%$, confirming that any potential information transfer is negligible. Together with the other tests, this means that the test set can be considered effectively independent.

%Training details.
Finally, we use the Adam optimizer \cite{adam2014method} with an initial learning rate $\eta$ = 1$^{-3}$, which anneals by 0.8 every 8 epochs. We implemented early stopping on the validation set with the patience set to 4 and training was complete after 24 epochs (see Fig. \ref{figx4}). Training was performed on an Nvidia 80 GB A100 hosted at the Jet Propulsion Laboratory and the run time was $\sim$ 3 hrs. We estimate that the total compute cost (cpu and gpu) across all experiments was roughly 150 hours.

%---------------------------------------------------------------------------------------
\subsection{Bayesian Inverse Modeling}
The model input parameters, $\boldsymbol{\theta}$, include the upstream Jovian magnetospheric O$^+$ density ($\rho_0$), Europa's atmospheric surface density ($n_0$), and Europa's atmospheric scale height ($H_0$). We represent the predicted magnetic field along a flyby using our transformer-based emulator, denoted by $f(\boldsymbol{\theta}) \in \mathbb{R}^3$, which outputs the mean of the magnetic field vector for a given set of input parameters. Given these predictions, we model the simulated MHD magnetic field $\mathbf{B}$ as normally distributed around our emulation, $f(\boldsymbol{\theta})$, with some assumed variance $\sigma^2$:

\begin{equation}
    p(\mathbf{B}|\boldsymbol{\theta}) = \mathcal{N}(f(\boldsymbol{\theta}), \sigma^2),
    \label{eq:likelihood}
\end{equation}
\\

In our probability model of the transformer-generated magnetic field, we include uncertainties as estimated from the emulation of the MHD model itself. We do this by assuming that the MAE errors reported in the results are a lower bound representation of the variance of the normal distribution of $p(\mathbf{B}|\boldsymbol{\theta})$. Specifically, we use $\sigma^2_{B_x} = \sigma^2_{B_y} = \sigma^2_{B_z} = 20$ nT$^2$ with covariance between magnetic field components as 0 (e.g. $\sigma^2_{B_x, B_y}=0$). This allows for the inclusion of an estimated uncertainty stemming from the emulation process itself into our parameter estimation. We specifically overestimate the uncertainty as our inversion (discussed below) includes samples from outside our training and test parameter space, and to illustrate the feasibility for highly uncertain observations. Future implementations of this approach can further constrain an estimation of $\sigma$ when performed on true observations. Note we do not include any observational uncertainties within $\sigma^2$ as our comparison is for hypothetical magnetic fields from the MHD model itself and thus including true observational uncertainties is not valid within this formulation.
Our interest within this section however, is not the forward model (the likelihood in Eq. \ref{eq:likelihood}), but the inverse. In other words, we aim to estimate the posterior \(p(\boldsymbol{\theta}|\mathbf{B})\). We connect the forward transformer, or likelihood model, to the posterior using Bayes' theorem \cite{Bayes1763, Kruschke2015}:

\begin{equation}
    p(\boldsymbol{\theta}|\mathbf{B}) = \frac{p(\mathbf{B}|\boldsymbol{\theta}) \; p(\boldsymbol{\theta})}{p(\mathbf{B})}.
    \label{eq:Bayes}
\end{equation}
\\

We assume normally distributed priors for $\boldsymbol{\theta}$ centered on the mean and standard deviation of the transformer’s training normalization. Calculations are performed in this normalized space and converted back to physical (and interpretable) values after estimation. All model input parameter priors are truncated at 0, which represents a nonphysical value for every input. For example a density ($\rho_0$) of less than zero is an input with zero percent probability. 
This approach does allow for parameter estimation outside the range of training of the emulator, which should be considered when interpreting the outputs of LEAP. In the case of parameter estimation outside of the training range, we recommend additional true MHD model runs to verify these results. 
The inversion was performed via \texttt{PyMC}, a probabilistic programming library in Python \cite{Salvatier2016, pymc}. The results presented herein were calculated with the slice sampler implementation within \texttt{PyMC}. The posterior is estimated using four chains, each with at least 500 emulated model evaluations. We discard the first 100 samples of each chain as burn-in, yielding a minimum of 1,600 posterior samples. We set the random seed to 42 for reproducibility.
Due to the nature of the likelihood, future interactions of LEAP may include gradient-informed Markov-chain Monte Carlo methods to better estimate the plasma parameters \cite[e.g.][]{NUTS}. 
\\*

%---------------------------------------------------------------------------------------
%\subsection{Parameter Importance}
%\lipsum[1-5]
%This is currently not a sedction, but it could be if we do something jazzier with it. E.g. really probe what is important and what is not.

%---------------------------------------------------------------------------------------
%---------------------------------------------------------------------------------------
%---------------------------------------------------------------------------------------
%---------------------------------------------------------------------------------------
%---------------------------------------------------------------------------------------
%---------------------------------------------------------------------------------------
%---------------------------------------------------------------------------------------
%---------------------------------------------------------------------------------------

\backmatter
\bmhead{Supplementary information}

\setcounter{figure}{0}
\setcounter{table}{0}
\renewcommand{\thefigure}{S\arabic{figure}}
\renewcommand{\thetable}{S\arabic{table}}

%EF 1. All flybys extracted and their positions wrt to Jupiter
\begin{figure}[h]
    \centering
    \includegraphics[width=1\textwidth]{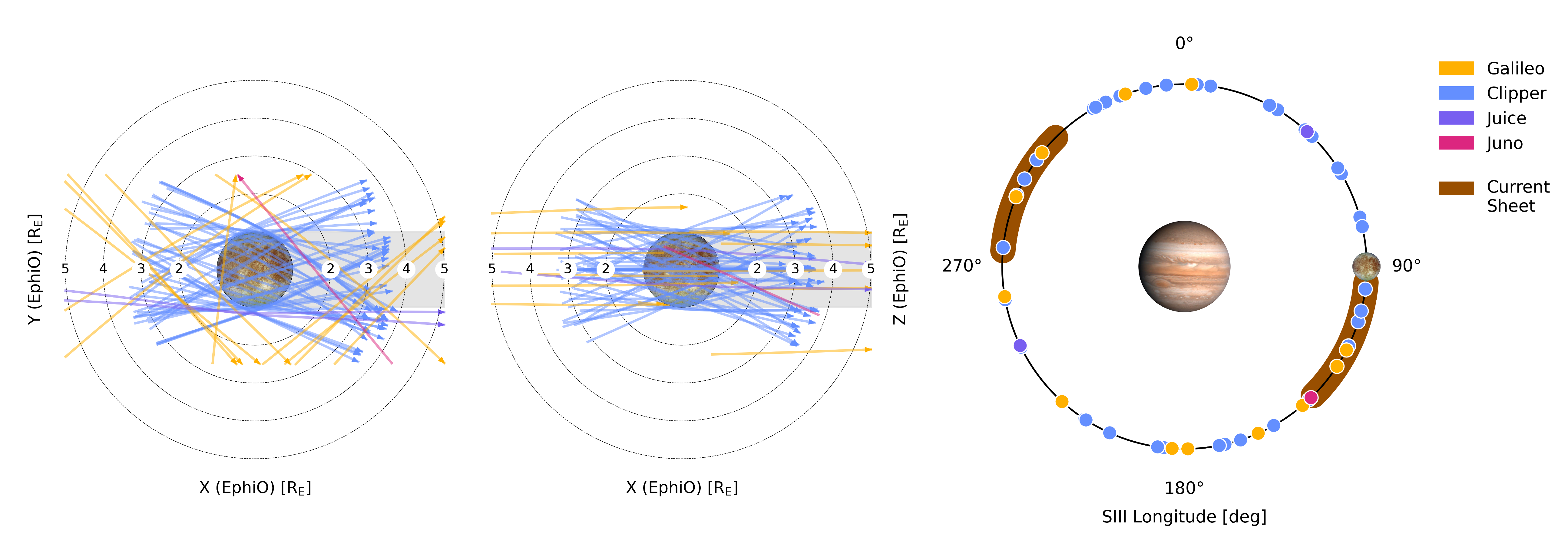}
    \caption{\textbf{All the flybys of Europa by the Galileo, Juno, Europa Clipper and JUICE missions with respect to the moon and Jovian current sheet. Data are in the EPhiO and SIII longitude coordinate system}}
    \label{figx1}
\end{figure}

\begin{figure}[h]
    \centering
    \includegraphics[width=1\textwidth]{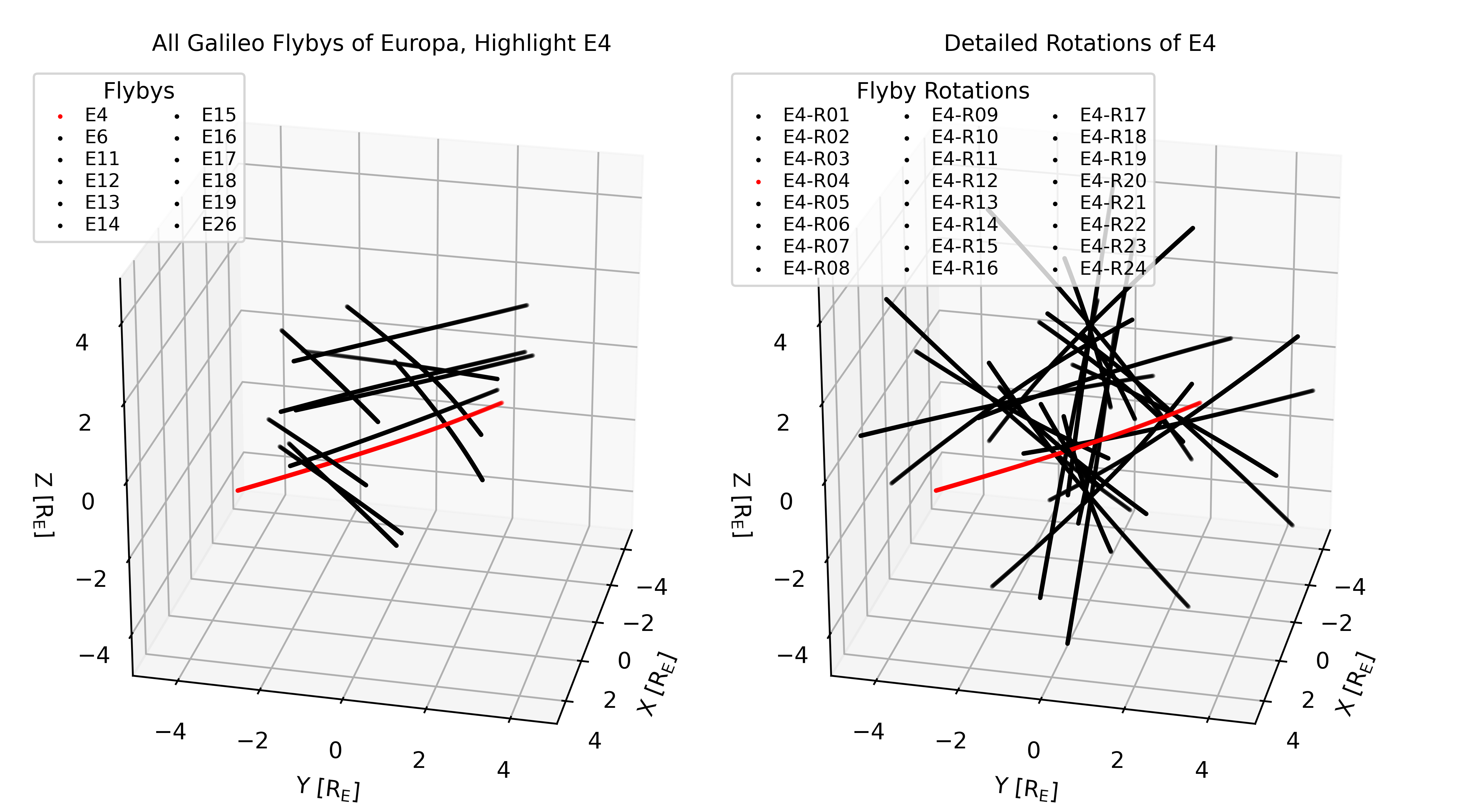}
    \caption{\textbf{Galileo $E4$ flyby extraction and 24 rotations}}
    \label{figx2}
\end{figure}

\begin{figure}[h]
    \centering
    \includegraphics[width=1\textwidth]{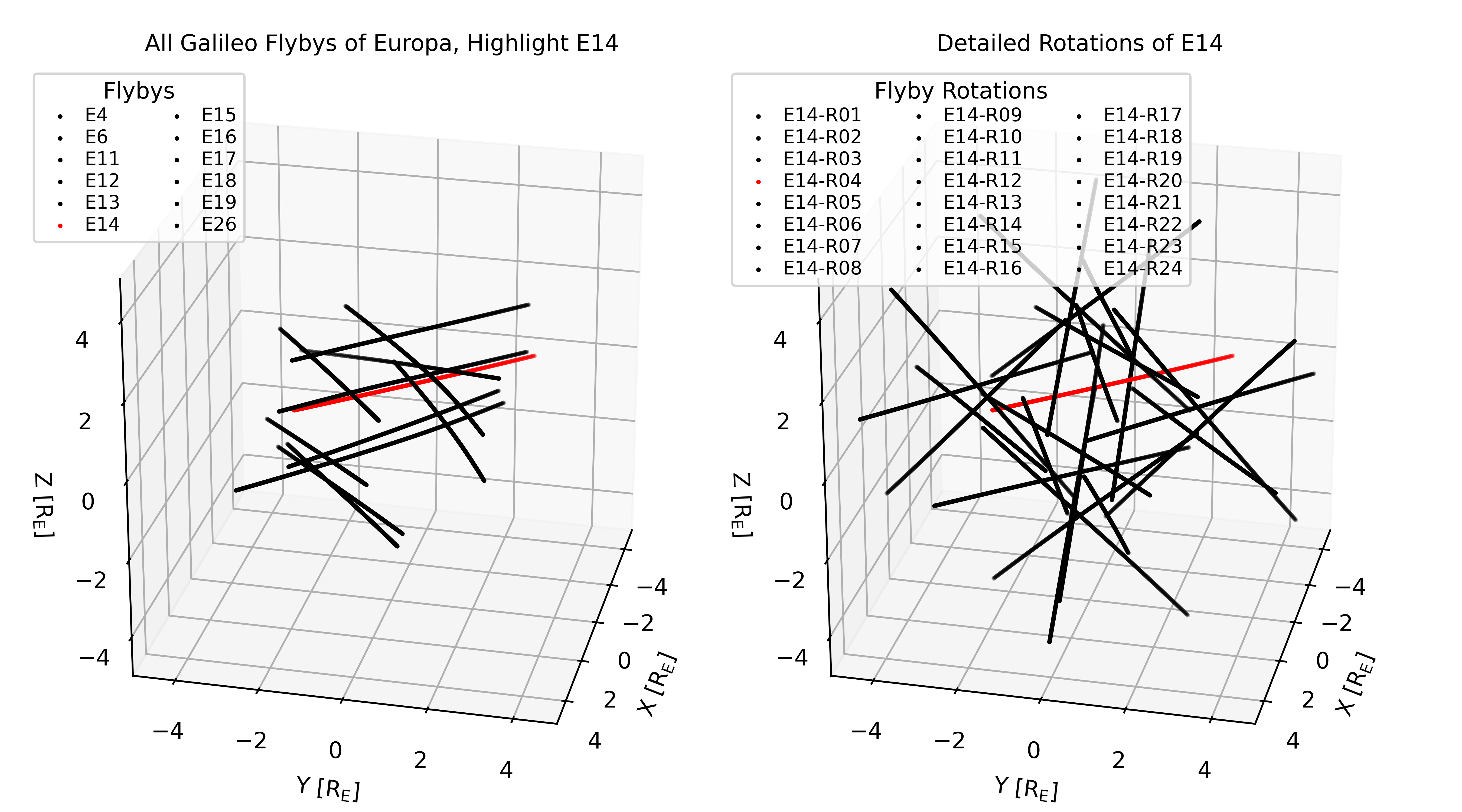}
    \caption{\textbf{Galileo $E14$ flyby extraction and 24 rotations}}
    \label{figx3}
\end{figure}

Figure \ref{figx1} illustrates all past and upcoming spacecraft flybys of Europa with respect to the moon, Jupiter, and its current sheet. Data are in the EPhiO and SIII coordinate system, with information extracted from the \texttt{NAIF SPICE} database and the \texttt{SpiceyPy} Python package. For each 3D volume we carve out these flybys and these are then rotated 24 times (Eq. \ref{eq:cube_rot_examples}). An example of this process applied to the Galileo $E4$ and $E14$ flybys shown in Figures \ref{figx2} and \ref{figx3}, respectively.

%EF 2-3. The $E4$ and $E14$ flybys and their rotations, as extract from the training set

\begin{figure}[h]
    \centering
    \includegraphics[width=0.6\textwidth]{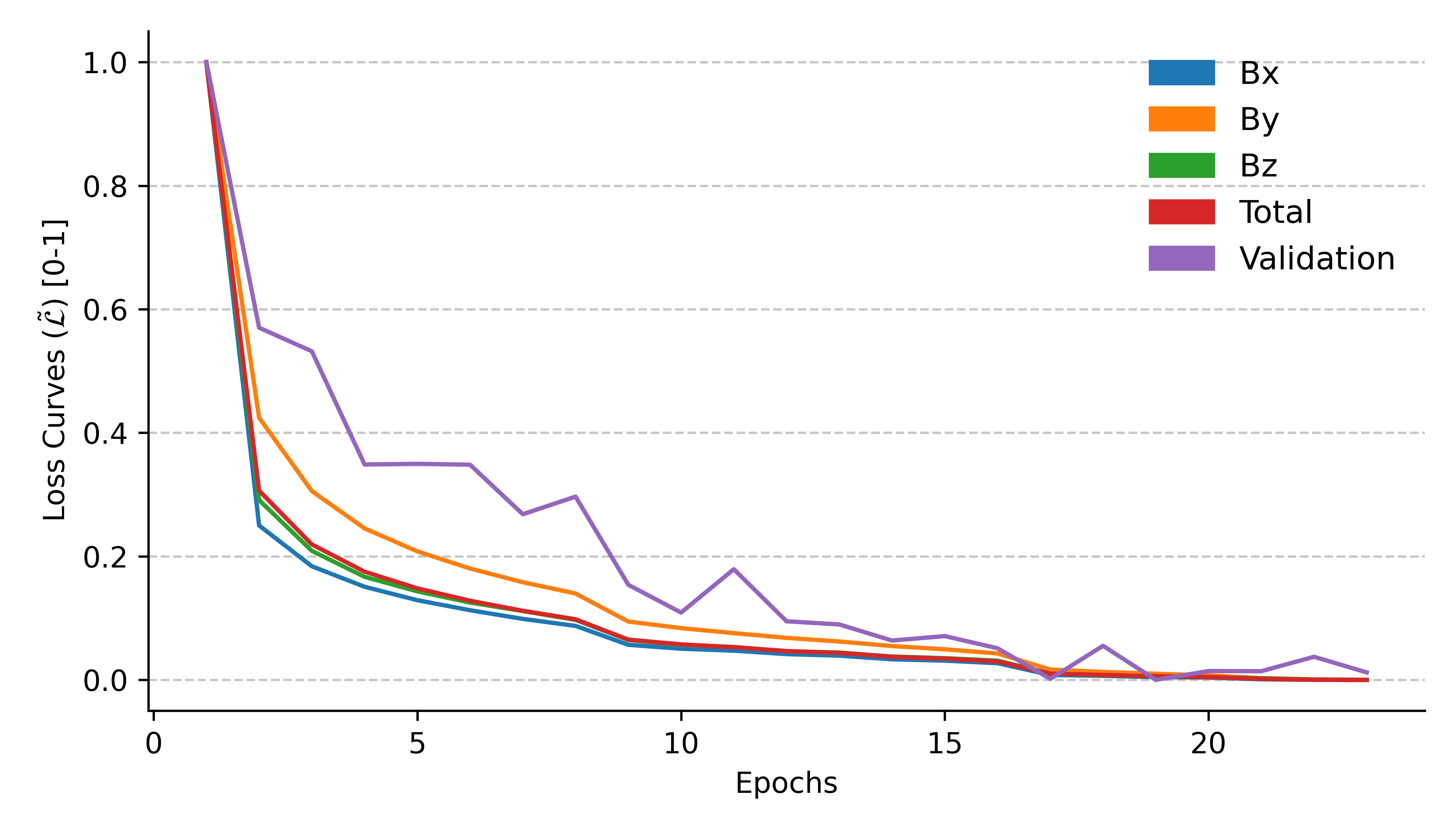}
    \caption{\textbf{Loss curves on the training set of the three magnetic field components ($B_x$, $B_y$, $B_z$) and total loss, and validation loss. }}
    \label{figx4}
\end{figure}

%EF5. The train and val curves
Figure \ref{figx4} shows the training and validation loss curves for the three magnetic field components. The training loss stops after the validation loss did not improve for 4 epochs i.e. 24 runs. The saw-tooth effect shown in the validation curve could be attributed to the limited training samples or hyperparameter settings, such as the patience. Since model selection is based on test set and acceptable \texttt{KFold} results, the appearance of these oscillations is considered cosmetic-only but may be examined further in future work.

%EF5. The train and val curves

% Table 3: Information for the Bayesian inference model from Fig 5
%ABBY
%You can see what i've shared in the other tables, so perhaps use these as a template and match with your sharing comfort levels :) - we can put in everything. Let me consider how to organize...

\begin{table}[h]
    \caption{Bayesian inversion posterior estimates on transformer based emulation as compared to true value. Results are visualized in Figure \ref{fig5}.}\label{tab3}
    \begin{tabular*}{\textwidth}{@{\extracolsep\fill}lcccc}
        \toprule
        & \multicolumn{4}{c}{True Values\footnotemark[1]} \\
        \cmidrule(lr){2-5}
        & $\rho_0$ [amu/cm$^{-3}$] & $P_0$ [nPa]  & $n_0$ [cm$^{-3}$, 10$^6$] & $H_0$ [km] \\
        \midrule
        All Flybys  & 320 & 0.414 & 25 & 100 \\
        & \multicolumn{4}{c}{Priors\footnotemark[2]} \\
        \cmidrule(lr){2-5}
        & $\rho_0$ [amu/cm$^{-3}$] & $P_0$ [nPa]  & $n_0$ [cm$^{-3}$, 10$^6$] & $H_0$ [km]\\
        \midrule
        All Flybys  & 1075$\pm$693 & 1.391$\pm$0.896 & 39.34$\pm$19.80 & 131.2$\pm$99.97  \\
        & \multicolumn{4}{c}{Posteriors\footnotemark[3]} \\
        \cmidrule(lr){2-5}
        & $\rho_0$ [amu/cm$^{-3}$] & $P_0$ [nPa]  & $n_0$ [cm$^{-3}$, 10$^6$] & $H_0$ [km] \\
        \midrule
        $E4$    & 252.1$\pm$137.7 & 0.616$\pm$0.238 & 27.44$\pm$4.218 & 96.37$\pm$5.89  \\
        $E14$   & 279.1$\pm$179.2 & 0.611$\pm$0.357 & 33.62$\pm$13.37 & 85.37$\pm$19.99  \\
        \botrule
    \end{tabular*}
    \footnotetext[1]{Known input value, expected to be roughly recovered by inversion.}
    \footnotetext[2]{All priors are assumed to be normally distributed and truncated at 0.}
    \footnotetext[3]{Range on values provided for priors and posteriors are the standard deviation. Note both of these are not exact normal distributions as they are truncated (prior) or sample based estimates (posterior).}
\end{table}
% Table 4: Information in GPU/HARDWARE, training time, floating points, size of model etc.

%---------------------------------------------------------------------------------------
%---------------------------------------------------------------------------------------
%---------------------------------------------------------------------------------------
%---------------------------------------------------------------------------------------
%---------------------------------------------------------------------------------------
%---------------------------------------------------------------------------------------
%---------------------------------------------------------------------------------------
%---------------------------------------------------------------------------------------

\bmhead{Acknowledgments}
%Acknowledgements are not compulsory. Where included they should be brief. Grant or contribution numbers may be acknowledged.%Please refer to Journal-level guidance for any specific requirements.

We thank Tobias Pfaff, Derek Posselt, Umaa Rebbapragada for constructive discussions regarding ML frameworks, modeling and sensitivity analyses. We acknowledge Al Cangahuala, Marc Costa and Arnaud Boutonnet for information related to the tour designs of Europa Clipper and JUICE. We also thank JPL's Center for Academic Partnerships which funded a technical exchange between JPL and the Halicioglu Data Science Institute (HDSI). We also acknowledge the Europa Clipper Magnetometer (ECM) and Plasma Instrument for Magnetic Sounding (PIMS) science teams for fruitful discussions and useful feedback. Finally, we thank the Jet Propulsion Laboratory High Performance Computing team for supporting this study. 

%---------------------------------------------------------------------------------------
%---------------------------------------------------------------------------------------
%---------------------------------------------------------------------------------------
%---------------------------------------------------------------------------------------
%---------------------------------------------------------------------------------------
%---------------------------------------------------------------------------------------
%---------------------------------------------------------------------------------------
%---------------------------------------------------------------------------------------

\section*{Declarations}
%Some journals require declarations to be submitted in a standardized format. Please check the Instructions for Authors of the journal to which you are submitting to see if you need to complete this section. If yes, your manuscript must contain the following sections under the heading `Declarations':

\begin{itemize}
\item Funding Declaration \newline
SR is supported by an appointment to the NASA Postdoctoral Program at the Jet Propulsion Laboratory, California Institute of Technology, administered by Oak Ridge Associated Universities (ORAU) through a contract with the National Aeronautics and Space Administration (NASA).
SR, CC, LM, and SV acknowledge support from the Jet Propulsion Laboratory, California Institute of Technology, under a contract with the National Aeronautics and Space Administration (contract 80NM0018D0004). AA acknowledges the support of the University of Alberta, as well as the Canada CIFAR AI Chairs program as hosted by the Alberta Machine Intelligence Institute. © 2026. All rights reserved.

\item Conflict of interest/Competing interests (check journal-specific guidelines for which heading to use) \newline
Not applicable

\item Ethics approval and consent to participate \newline
Not applicable

\item Consent for publication \newline
Yes

\item Data availability \newline
The MHD training, validation, and test sets are on Hugging Face (\url{https://huggingface.co/datasets/reddysachin/LEAP_dataset}). The pre-trained model is also on Hugging Face (\url{https://huggingface.co/reddysachin/LEAP}). The \textit{SPICE} kernels are freely available via the Planetary Data System Navigation Node (\url{https://naif.jpl.nasa.gov/pub/naif/}). The observational data is on the Planetary Data Science node (\url{https://pds.nasa.gov/}) and in the literature \cite{kabin1999europa, gurnett1998galileo, harris2021multi, harris2022multi}.

\item Materials availability \newline
Not applicable

\item Code availability \newline
The code repository for this research is available on GitHub (\url{https://github.com/reddy-sachin/LEAP/tree/main}) [note this is private until after publication. Reviewers have access to a zip file]. The BATSRUS code is available on GitHub (\url{https://github.com/SWMFsoftware/BATSRUS}) and is in the literature \cite{harris2021multi, harris2022multi, nordheim2022magnetospheric}.

\item Author contribution \newline
SR conceived and led the study, built the transformer, extracted the flybys, performed the parameter estimation/survey and ablation test, and co-wrote the manuscript. AA created the Bayesian inversion schema and co-wrote the manuscript. CC co-managed the study and extracted the flybys. XJ co-managed the study and, XJ and CH built the MHD code and provided the training data. TN co-managed the study and provided additional training data. LM co-designed the data leakage experiments. CC, XJ, and SV informed and validated the ocean induction discussion. IC validated the transformer framework. All authors analyzed the results and contributed to the editing of the manuscript. © 2026. All rights reserved.

\item Competing Interest \newline
The authors declare no competing interests.

\end{itemize}

\noindent

\bibliography{sn-bibliography}% common bib file

@article{harris2021multi,
  title={Multi-fluid MHD simulations of Europa's plasma interaction under different magnetospheric conditions},
  author={Harris, Camilla DK and Jia, Xianzhe and Slavin, James A and Toth, Gabor and Huang, Zhenguang and Rubin, Martin},
  journal={Journal of Geophysical Research: Space Physics},
  volume={126},
  number={5},
  pages={e2020JA028888},
  year={2021},
  publisher={Wiley Online Library}
}

@article{harris2022multi,
  title={Multi-fluid MHD simulations of Europa's plasma interaction: Effects of variation in Europa's atmosphere},
  author={Harris, Camilla DK and Jia, Xianzhe and Slavin, James A},
  journal={Journal of Geophysical Research: Space Physics},
  volume={127},
  number={9},
  pages={e2022JA030569},
  year={2022},
  publisher={Wiley Online Library}
}

@article{vaswani2017attention,
  title={Attention is all you need},
  author={Vaswani, Ashish and Shazeer, Noam and Parmar, Niki and Uszkoreit, Jakob and Jones, Llion and Gomez, Aidan N and Kaiser, {\L}ukasz and Polosukhin, Illia},
  journal={Advances in neural information processing systems},
  volume={30},
  year={2017}
}

@article{rubin2015self,
  title={Self-consistent multifluid MHD simulations of Europa's exospheric interaction with Jupiter's magnetosphere},
  author={Rubin, M and Jia, X and Altwegg, K and Combi, MR and Daldorff, LKS and Gombosi, TI and Khurana, K and Kivelson, MG and Tenishev, VM and T{\'o}th, G and others},
  journal={Journal of Geophysical Research: Space Physics},
  volume={120},
  number={5},
  pages={3503--3524},
  year={2015},
  publisher={Wiley Online Library}
}

@article{kivelson2004magnetospheric,
  title={Magnetospheric interactions with satellites},
  author={Kivelson, Margaret G and Bagenal, Fran and Kurth, William S and Neubauer, Fritz M and Paranicas, Chris and Saur, Joachim},
  journal={Jupiter: The planet, satellites and magnetosphere},
  volume={1},
  pages={513--536},
  year={2004}
}

@article{glocer2009multifluid,
  title={Multifluid block-adaptive-tree solar wind roe-type upwind scheme: Magnetospheric composition and dynamics during geomagnetic storms—Initial results},
  author={Glocer, A and T{\'o}th, G and Ma, Y and Gombosi, T and Zhang, J-C and Kistler, LM},
  journal={Journal of Geophysical Research: Space Physics},
  volume={114},
  number={A12},
  year={2009},
  publisher={Wiley Online Library}
}

@article{toth2012adaptive,
  title={Adaptive numerical algorithms in space weather modeling},
  author={T{\'o}th, G{\'a}bor and Van der Holst, Bart and Sokolov, Igor V and De Zeeuw, Darren L and Gombosi, Tamas I and Fang, Fang and Manchester, Ward B and Meng, Xing and Najib, Dalal and Powell, Kenneth G and others},
  journal={Journal of Computational Physics},
  volume={231},
  number={3},
  pages={870--903},
  year={2012},
  publisher={Elsevier}
}

@article{plainaki2013exospheric,
  title={Exospheric O2 densities at Europa during different orbital phases},
  author={Plainaki, C and Milillo, A and Mura, A and Saur, J and Orsini, S and Massetti, S},
  journal={Planetary and Space Science},
  volume={88},
  pages={42--52},
  year={2013},
  publisher={Elsevier}
}

@article{kliore1997ionosphere,
  title={The ionosphere of Europa from Galileo radio occultations},
  author={Kliore, Arvydas J and Hinson, David P and Flasar, F Michael and Nagy, Andrew F and Cravens, Thomas E},
  journal={Science},
  volume={277},
  number={5324},
  pages={355--358},
  year={1997},
  publisher={American Association for the Advancement of Science}
}

@article{cangahuala2025europa,
  title={Europa Clipper mission design, mission plan, and navigation},
  author={Cangahuala, L Alberto and Campagnola, Stefano and Bradley, Ben K and Boone, Dylan R and Buffington, Brent B and Ludwinski, Jan M and Nandi, Sumita and Scott, Christopher J},
  journal={Space Science Reviews},
  volume={221},
  number={1},
  pages={22},
  year={2025},
  publisher={Springer}
}

@article{boutonnet2024designing,
  title={Designing the JUICE trajectory},
  author={Boutonnet, A and Langevin, Y and Erd, C},
  journal={Space Science Reviews},
  volume={220},
  number={6},
  pages={67},
  year={2024},
  publisher={Springer}
}

@article{kivelson2023ecm,
  title={The Europa clipper magnetometer},
  author={Kivelson, Margaret G and Jia, Xianzhe and Lee, Karen A and Raymond, Carol A and Khurana, Krishan K and Perley, Mitchell O and Biersteker, John B and Blacksberg, Jordana and Caron, Ryan and Cochrane, Corey J and others},
  journal={Space Science Reviews},
  volume={219},
  number={6},
  pages={48},
  year={2023},
  publisher={Springer}
}

@article{vance2023investigating,
  title={Investigating Europa’s habitability with the Europa Clipper},
  author={Vance, Steven D and Craft, Kathleen L and Shock, Everett and Schmidt, Britney E and Lunine, Jonathan and Hand, Kevin P and McKinnon, William B and Spiers, Elizabeth M and Chivers, Chase and Lawrence, Justin D and others},
  journal={Space Science Reviews},
  volume={219},
  number={8},
  pages={81},
  year={2023},
  publisher={Springer}
}

@article{pappalardo2024science,
  title={Science overview of the Europa clipper mission},
  author={Pappalardo, Robert T and Buratti, Bonnie J and Korth, Haje and Senske, David A and Blaney, Diana L and Blankenship, Donald D and Burch, James L and Christensen, Philip R and Kempf, Sascha and Kivelson, Margaret G and others},
  journal={Space Science Reviews},
  volume={220},
  number={4},
  pages={40},
  year={2024},
  publisher={Springer}
}

@article{cervantes2025mhd,
  title={MHD simulations of Europa's interaction with Jupiter's magnetosphere during the Juno flyby: Electron beams in the plasma wake},
  author={Cervantes, Sebastian and Saur, Joachim and Duling, Stefan and Szalay, Jamey R and Schlegel, Stephan and Connerney, John EP and Allegrini, Frederic and Bolton, S},
  journal={Journal of Geophysical Research: Space Physics},
  volume={130},
  number={6},
  pages={e2025JA033825},
  year={2025},
  publisher={Wiley Online Library}
}

@article{addison2024magnetic,
  title={Magnetic signatures of the interaction between Europa and Jupiter's magnetosphere during the Juno flyby},
  author={Addison, Peter and Haynes, C Michael and Stahl, Aaron M and Liuzzo, Lucas and Simon, Sven},
  journal={Geophysical Research Letters},
  volume={51},
  number={2},
  pages={e2023GL106810},
  year={2024},
  publisher={Wiley Online Library}
}

@article{allegrini2024electron,
  title={Electron beams at Europa},
  author={Allegrini, F and Saur, J and Szalay, JR and Ebert, RW and Kurth, WS and Cervantes, S and Smith, HT and Bagenal, F and Bolton, SJ and Clark, G and others},
  journal={Geophysical research letters},
  volume={51},
  number={13},
  pages={e2024GL108422},
  year={2024},
  publisher={Wiley Online Library}
}

@article{neubauer1999alfven,
  title={Alfv{\'e}n wings and electromagnetic induction in the interiors: Europa and Callisto},
  author={Neubauer, Fritz M},
  journal={Journal of Geophysical Research: Space Physics},
  volume={104},
  number={A12},
  pages={28671--28684},
  year={1999},
  publisher={Wiley Online Library}
}

@article{volwerk2007europa,
  title={Europa's Alfv{\'e}n wing: Shrinkage and displacement influenced by an induced magnetic field},
  author={Volwerk, M and Khurana, K and Kivelson, M},
  journal={Annales Geophysicae},
  volume={25},
  number={4},
  pages={905--914},
  year={2007},
  organization={Copernicus Publications G{\"o}ttingen, Germany}
}

@article{kabin1999europa,
  title={On Europa's magnetospheric interaction: A MHD simulation of the E4 flyby},
  author={Kabin, K and Combi, MR and Gombosi, TI and Nagy, AF and DeZeeuw, DL and Powell, KG},
  journal={Journal of Geophysical Research: Space Physics},
  volume={104},
  number={A9},
  pages={19983--19992},
  year={1999},
  publisher={Wiley Online Library}
}

@article{gurnett1998galileo,
  title={Galileo plasma wave observations near Europa},
  author={Gurnett, DA and Kurth, WS and Roux, A and Bolton, SJ and Thomsen, EA and Groene, JB},
  journal={Geophysical research letters},
  volume={25},
  number={3},
  pages={237--240},
  year={1998},
  publisher={Wiley Online Library}
}

@article{kivelson2000galileo,
  title={Galileo magnetometer measurements: A stronger case for a subsurface ocean at Europa},
  author={Kivelson, Margaret G and Khurana, Krishan K and Russell, Christopher T and Volwerk, Martin and Walker, Raymond J and Zimmer, Christophe},
  journal={Science},
  volume={289},
  number={5483},
  pages={1340--1343},
  year={2000},
  publisher={American Association for the Advancement of Science}
}

@article{khurana1998induced,
  title={Induced magnetic fields as evidence for subsurface oceans in Europa and Callisto},
  author={Khurana, KK and Kivelson, MG and Stevenson, DJ and Schubert, G and Russell, CT and Walker, RJ and Polanskey, C},
  journal={Nature},
  volume={395},
  number={6704},
  pages={777--780},
  year={1998},
  publisher={Nature Publishing Group UK London}
}

@article{zimmer2000subsurface,
  title={Subsurface oceans on Europa and Callisto: Constraints from Galileo magnetometer observations},
  author={Zimmer, Christophe and Khurana, Krishan K and Kivelson, Margaret G},
  journal={Icarus},
  volume={147},
  number={2},
  pages={329--347},
  year={2000},
  publisher={Elsevier}
}

@article{jia2010magnetic,
  title={Magnetic fields of the satellites of Jupiter and Saturn},
  author={Jia, Xianzhe and Kivelson, Margaret G and Khurana, Krishan K and Walker, Raymond J},
  journal={Space Science Reviews},
  volume={152},
  number={1},
  pages={271--305},
  year={2010},
  publisher={Springer}
}

@article{tanaka1994finite,
  title={Finite volume TVD scheme on an unstructured grid system for three-dimensional MHD simulation of inhomogeneous systems including strong background potential fields},
  author={Tanaka, T},
  journal={Journal of Computational Physics},
  volume={111},
  number={2},
  pages={381--389},
  year={1994},
  publisher={Elsevier}
}

@article{kataoka2024machine,
  title={Machine Learning-Based Emulator for the Physics-Based Simulation of Auroral Current System},
  author={Kataoka, Ryuho and Nakamizo, Aoi and Nakano, Shinya and Fujita, Shigeru},
  journal={Space Weather},
  volume={22},
  number={1},
  pages={e2023SW003720},
  year={2024},
  publisher={Wiley Online Library}
}

@article{schilling2008influence,
  title={Influence of the internally induced magnetic field on the plasma interaction of Europa},
  author={Schilling, N and Neubauer, FM and Saur, J},
  journal={Journal of Geophysical Research: Space Physics},
  volume={113},
  number={A3},
  year={2008},
  publisher={Wiley Online Library}
}

@article{saur1998interaction,
  title={Interaction of the Jovian magnetosphere with Europa: Constraints on the neutral atmosphere},
  author={Saur, J and Strobel, DF and Neubauer, FM},
  journal={Journal of Geophysical Research: Planets},
  volume={103},
  number={E9},
  pages={19947--19962},
  year={1998},
  publisher={Wiley Online Library}
}

@article{jia2018evidence,
  title={Evidence of a plume on Europa from Galileo magnetic and plasma wave signatures},
  author={Jia, Xianzhe and Kivelson, Margaret G and Khurana, Krishan K and Kurth, William S},
  journal={Nature Astronomy},
  volume={2},
  number={6},
  pages={459--464},
  year={2018},
  publisher={Nature Publishing Group UK London}
}

@article{muller2011aikef,
  title={AIKEF: Adaptive hybrid model for space plasma simulations},
  author={M{\"u}ller, Joachim and Simon, Sven and Motschmann, Uwe and Sch{\"u}le, Josef and Glassmeier, Karl-Heinz and Pringle, Gavin J},
  journal={Computer Physics Communications},
  volume={182},
  number={4},
  pages={946--966},
  year={2011},
  publisher={Elsevier}
}

@article{cochrane2025stronger,
  title={Stronger evidence of a subsurface ocean within Callisto from a multifrequency investigation of Its induced magnetic field},
  author={Cochrane, Corey J and Vance, Steven D and Castillo-Rogez, Julie C and Styczinski, Marshall J and Liuzzo, Lucas},
  journal={AGU Advances},
  volume={6},
  number={1},
  pages={e2024AV001237},
  year={2025},
  publisher={Wiley Online Library}
}

@misc{frazier2020trident,
  title     = {Trident: the path to Triton on a discovery budget},
  author    = {Frazier, William and Bearden, David and Mitchell, Karl L. and Lam, Try and Prockter, Louise and Dissly, Richard},
  booktitle = {IEEE Aerospace Conference},
  year      = {2020},
  pages     = {1--12},
  publisher = {IEEE}
}

@article{girija2023flagship,
  title={A Flagship-class Uranus Orbiter and Probe mission concept using aerocapture},
  author={Girija, Athul Pradeepkumar},
  journal={Acta Astronautica},
  volume={202},
  pages={104--118},
  year={2023},
  publisher={Elsevier}
}

@article{oza2019dusk,
  title={Dusk over dawn O2 asymmetry in Europa's near-surface atmosphere},
  author={Oza, Apurva V and Leblanc, Francois and Johnson, Robert E and Schmidt, Carl and Leclercq, Ludivine and Cassidy, Timothy A and Chaufray, Jean-Yves},
  journal={Planetary and Space Science},
  volume={167},
  pages={23--32},
  year={2019},
  publisher={Elsevier}
}

@article{neubauer1980nonlinear,
  title={Nonlinear standing Alfv{\'e}n wave current system at Io: Theory},
  author={Neubauer, FM},
  journal={Journal of Geophysical Research: Space Physics},
  volume={85},
  number={A3},
  pages={1171--1178},
  year={1980},
  publisher={Wiley Online Library}
}

@article{Grasset2013,
  title={JUpiter ICy moons Explorer (JUICE): An ESA mission to orbit Ganymede and to characterise the Jupiter system},
  author={Grasset, Olivier and Dougherty, MK and Coustenis, A and Bunce, EJ and Erd, C and Titov, D and Blanc, M and Coates, A and Drossart, P and Fletcher, LN and others},
  journal={Planetary and Space Science},
  volume={78},
  pages={1--21},
  year={2013},
  publisher={Elsevier}
}

@article{nordheim2022magnetospheric,
  title={Magnetospheric ion bombardment of Europa’s surface},
  author={Nordheim, TA and Regoli, LH and Harris, CDK and Paranicas, Christopher and Hand, KP and Jia, Xianzhe},
  journal={The Planetary Science Journal},
  volume={3},
  number={1},
  pages={5},
  year={2022},
  publisher={IOP Publishing}
}

@article{burkhart2020catalogue,
  title={The catalogue for astrophysical turbulence simulations (cats)},
  author={Burkhart, B and Appel, SM and Bialy, S and Cho, J and Christensen, AJ and Collins, D and Federrath, Christoph and Fielding, DB and Finkbeiner, D and Hill, AS and others},
  journal={The Astrophysical Journal},
  volume={905},
  number={1},
  pages={14},
  year={2020},
  publisher={IOP Publishing}
}

@article{westlake2023plasma,
  title={The Plasma Instrument for Magnetic Sounding (PIMS) on the Europa Clipper Mission},
  author={Westlake, Joseph H and McNutt Jr, RL and Grey, M and Coren, D and Rymer, AM and Cochrane, CJ and Luspay-Kuti, A and Hohlfeld, E and Seese, N and Crew, A and others},
  journal={Space Science Reviews},
  volume={219},
  number={8},
  pages={62},
  year={2023},
  publisher={Springer}
}

@article{
Cranmer2020,
author = {Kyle Cranmer  and Johann Brehmer  and Gilles Louppe },
title = {The frontier of simulation-based inference},
journal = {Proceedings of the National Academy of Sciences},
volume = {117},
number = {48},
pages = {30055-30062},
year = {2020},
doi = {10.1073/pnas.1912789117},
URL = {https://www.pnas.org/doi/abs/10.1073/pnas.1912789117},
eprint = {https://www.pnas.org/doi/pdf/10.1073/pnas.1912789117}}

@article{Salvatier2016,
author = {Salvatier, J. and Wiecki, T.V. and Fonnesbeck, C.},
title = {Probabilistic programming in {Python} using {PyMC3}},
pages = {2:e55},
journal = {PeerJ Computer Science},
doi = {10.7717/peerj-cs.55},
year = {2016}
}

@article{pymc,
  author = {Abril-Pla, Oriol and Andreani, Virgile and Carroll, Colin and Dong, Larry and Fonnesbeck, Christopher J. and Kochurov, Maxim and Kumar, Ravin and Lao, Junpeng and Luhmann, Christian C. and Martin, Osvaldo A. and Osthege, Michael and Vieira, Ricardo and Wiecki, Thomas and Zinkov, Robert},
  title = {{PyMC: a modern, and comprehensive probabilistic programming framework in Python}},
  journal = {PeerJ Computer Science},
  volume = {9},
  pages = {e1516},
  year = {2023},
  publisher = {PeerJ Inc.},
  doi = {10.7717/peerj-cs.1516},
}

@article{NUTS,
  author = {Hoffman, M. D. and Gelman, A.},
  title = {{The No-U-Turn Sampler: Adaptively Setting Path Lengths in Hamiltonian Monte Carlo}},
  journal = {Journal of Machine Learning Research},
  volume = {15},
  pages = {1593-1623},
  year = {2011},
}

@article{Bayes1763,
author = {Bayes, T. and Price, R.},
title = {An Essay towards Solving a Problem in the Doctrine of Chances. {B}y the Late {Rev.} {Mr.} {Bayes}, {F.} {R.} {S.} Communicated by {Mr.} {Price}, in a Letter to {John} {Canton}, {A.} {M.} {F.} {R.} {S}},
volume = {53},
pages = {370–418},
journal = {Philosophical Transactions of the Royal Society of London},
doi = {10.1098/rstl.1763.0053},
year = {1763}
}

@book{Kruschke2015,
author = {Kruschke, John K.},
title = {Doing {Bayesian} Data Analysis: A Tutorial with {R}, {JAGS}, and {Stan}},
publisher = {Elsevier},
ed = {2nd},
isbn = {978-0-12-405888-0},
year = {2015}
}

@article{adam2014method,
  title={A method for stochastic optimization},
  author={Adam, Kingma DP Ba J and others},
  journal={arXiv preprint arXiv:1412.6980},
  volume={1412},
  number={6},
  year={2014}
}

@article{gombosi2021swmf,
author = {{Gombosi}, Tamas I. and {Chen}, Yuxi and {Glocer}, Alex and {Huang}, Zhenguang and {Jia}, Xianzhe and {Liemohn}, Michael W. and {Manchester}, Ward B. and {Pulkkinen}, Tuija and {Sachdeva}, Nishtha and {Al Shidi}, Qusai and {Sokolov}, Igor V. and {Szente}, Judit and {Tenishev}, Valeriy and {Toth}, Gabor and {van der Holst}, Bart and {Welling}, Daniel T. and {Zhao}, Lulu and {Zou}, Shasha},
        title = "{What sustained multi-disciplinary research can achieve: The space weather modeling framework}",
      journal = {Journal of Space Weather and Space Climate},
         year = 2021,
        month = may,
       volume = {11},
          eid = {42},
        pages = {42},
          doi = {10.1051/swsc/2021020}
}

@article{styczinski2023planetprofile,
  title={PlanetProfile: self-consistent interior structure modeling for ocean worlds and rocky dwarf planets in python},
  author={Styczinski, MJ and Vance, SD and Melwani Daswani, M},
  journal={Earth and Space Science},
  volume={10},
  number={8},
  pages={e2022EA002748},
  year={2023},
  publisher={Wiley Online Library}
}

@article{huang2020tabtransformer,
  title={Tabtransformer: Tabular data modeling using contextual embeddings},
  author={Huang, Xin and Khetan, Ashish and Cvitkovic, Milan and Karnin, Zohar},
  journal={arXiv preprint arXiv:2012.06678},
  year={2020}
}

@article{khurana1997euler,
  title={Euler potential models of Jupiter's magnetospheric field},
  author={Khurana, Krishan K},
  journal={Journal of Geophysical Research: Space Physics},
  volume={102},
  number={A6},
  pages={11295--11306},
  year={1997},
  publisher={Wiley Online Library}
}

@InProceedings{holzschuh2025transformer,
  title = 	 {{PDE}-Transformer: Efficient and Versatile Transformers for Physics Simulations},
  author =       {Holzschuh, Benjamin and Liu, Qiang and Kohl, Georg and Thuerey, Nils},
  booktitle = 	 {Proceedings of the 42nd International Conference on Machine Learning},
  pages = 	 {23562--23602},
  year = 	 {2025},
  editor = 	 {Singh, Aarti and Fazel, Maryam and Hsu, Daniel and Lacoste-Julien, Simon and Berkenkamp, Felix and Maharaj, Tegan and Wagstaff, Kiri and Zhu, Jerry},
  volume = 	 {267},
  series = 	 {Proceedings of Machine Learning Research},
  month = 	 {13--19 Jul},
  publisher =    {PMLR},
  url = 	 {https://proceedings.mlr.press/v267/holzschuh25a.html},
}

@misc{wan2025pesanet,
      title={PeSANet: Physics-encoded Spectral Attention Network for Simulating PDE-Governed Complex Systems}, 
      author={Han Wan and Rui Zhang and Qi Wang and Yang Liu and Hao Sun},
      year={2025},
      eprint={2505.01736},
      archivePrefix={arXiv},
      primaryClass={cs.LG},
      url={https://arxiv.org/abs/2505.01736}, 
}

@article{masters2025magnetosphere,
  title={Magnetosphere and plasma science with the Jupiter Icy Moons Explorer},
  author={Masters, Adam and Modolo, Ronan and Roussos, Elias and Krupp, Norbert and Witasse, Olivier and Vallat, Claire and Cecconi, Baptiste and Edberg, Niklas JT and Futaana, Yoshifumi and Galand, Marina and others},
  journal={Space Science Reviews},
  volume={221},
  number={2},
  pages={24},
  year={2025},
  publisher={Springer}
}

@misc{holmberg2025graph,
  title        = {Graph-based Neural Space Weather Forecasting},
  author       = {Holmberg, Daniel and Zaitsev, Ivan and Alho, Markku and Bouri, Ioanna and Franssila, Fanni and Jeong, Haewon and Palmroth, Minna and Roos, Teemu},
  booktitle    = {NeurIPS 2025 Workshop on Machine Learning and the Physical Sciences},
  pages        = {1--6},
  year         = {2025},
  organization = {NeurIPS}
}

@article{von2014vlasiator,
  title={Vlasiator: First global hybrid-Vlasov simulations of Earth's foreshock and magnetosheath},
  author={Von Alfthan, S and Pokhotelov, D and Kempf, Y and Hoilijoki, S and Honkonen, I and Sandroos, A and Palmroth, M},
  journal={Journal of Atmospheric and Solar-Terrestrial Physics},
  volume={120},
  pages={24--35},
  year={2014},
  publisher={Elsevier}
}

@misc{azari2023simulation,
  author       = {Azari, A. R. and Ermakov, A. and Harris, C. D. K. and 
                  Liuzzo, L. and Abrahams, E. and Duling, S.},
  title        = {Simulation Based Inference for Mission Planning and Parameter Estimation},
  booktitle    = {Uranus Flagship: Investigations and Instruments for Cross-Discipline Science Workshop},
  year         = {2023},
  volume       = {2808},
  pages        = {8069},
  organization = {Lunar and Planetary Institute}
}

@article{duling2022ganymede,
  title={Ganymede MHD model: Magnetospheric context for Juno's PJ34 flyby},
  author={Duling, Stefan and Saur, Joachim and Clark, George and Allegrini, Frederic and Greathouse, Thomas and Gladstone, Randy and Kurth, William and Connerney, John EP and Bagenal, Fran and Sulaiman, Ali H},
  journal={Geophysical Research Letters},
  volume={49},
  number={24},
  pages={e2022GL101688},
  year={2022},
  publisher={Wiley Online Library}
}

@article{cochrane2023magnetic,
  title={Magnetic field modeling and visualization of the Europa Clipper spacecraft},
  author={Cochrane, Corey J and Murphy, Neil and Raymond, Carol A and Biersteker, John B and Dang, Katherine and Jia, Xianzhe and Korth, Haje and Narvaez, Pablo and Ream, Jodie B and Weiss, Benjamin P},
  journal={Space Science Reviews},
  volume={219},
  number={4},
  pages={34},
  year={2023},
  publisher={Springer}
}

@article{saur2010induced,
  title={Induced magnetic fields in solar system bodies},
  author={Saur, Joachim and Neubauer, Fritz M and Glassmeier, Karl-Heinz},
  journal={Space science reviews},
  volume={152},
  number={1},
  pages={391--421},
  year={2010},
  publisher={Springer}
}

@article{cochrane2024detecting,
  title={On detecting and characterizing planetary oceans in the solar system using a distance-based ensemble modelling approach: application to the Uranus system},
  author={Cochrane, CJ and Vance, SD and Biersteker, JB and Styczinski, MJ and Weiss, B},
  journal={Philosophical Transactions A},
  volume={382},
  number={2286},
  pages={20240086},
  year={2024},
  publisher={The Royal Society}
}

@article{khurana2009electromagnetic,
  title={Electromagnetic induction from Europa’s ocean and the deep interior},
  author={Khurana, Krishan K and Kivelson, Margaret G and Hand, Kevin P and Russell, Christopher T},
  journal={Europa},
  pages={572--586},
  year={2009},
  publisher={University of Arizona Press Tucson, AZ}
}

@article{biersteker2023revealing,
  title={Revealing the interior structure of icy moons with a Bayesian approach to magnetic induction measurements},
  author={Biersteker, John B and Weiss, Benjamin P and Cochrane, Corey J and Harris, Camilla DK and Jia, Xianzhe and Khurana, Krishan K and Liu, Jiang and Murphy, Neil and Raymond, Carol A},
  journal={The Planetary Science Journal},
  volume={4},
  number={4},
  pages={62},
  year={2023},
  publisher={The American Astronomical Society}
}

@article{winkenstern2025inversion,
  title={An inversion of magnetic field measurements to constrain the depth, thickness, and conductivity of Europa's ocean},
  author={Winkenstern, Jason and Saur, Joachim and Cervantes, Sebastian},
  journal={Journal of Geophysical Research: Planets},
  volume={130},
  number={10},
  pages={e2025JE009122},
  year={2025},
  publisher={Wiley Online Library}
}

@article{styczinski2024planetmag,
  title={PlanetMag: Software for evaluation of outer planet magnetic fields and corresponding excitations at their moons},
  author={Styczinski, Marshall J and Cochrane, Corey Jonathan},
  journal={Earth and Space Science},
  volume={11},
  number={6},
  pages={e2024EA003552},
  year={2024},
  publisher={Wiley Online Library}
}

@article{meitzler2023investigating,
  title={Investigating Europa’s radiation environment with the Europa Clipper radiation monitor},
  author={Meitzler, Richard and Jun, Insoo and Blase, Ryan and Cassidy, Timothy and Clark, Roger and Cochrane, Corey and Fix, Sam and Gladstone, Randy and Goldsten, John and Gudipati, Murthy and others},
  journal={Space Science Reviews},
  volume={219},
  number={7},
  pages={61},
  year={2023},
  publisher={Springer}
}

%TC:endignore
\end{document}